\documentclass[aps,prb,showpacs,superscriptaddress]{revtex4}

\usepackage{graphicx,amssymb,amsfonts,amsmath}

\def \onlinesquare#1{\onlinecite{#1}}

\def \bw{{}}

\def \tR{{\tilde{R}}}

\usepackage[latin1]{inputenc}
\usepackage{amsmath}
\usepackage{amsfonts}
\usepackage{amssymb}
\usepackage{graphicx}
\usepackage[usenames]{color}
\usepackage{epstopdf}

\newcommand{\be}{\begin{equation}}
\newcommand{\ee}{\end{equation}}

\def \be{\begin{equation}}
\def \ee{\end{equation}}
\def \ba{\begin{array}}
\def \ea{\end{array}}
\def \bea{\begin{eqnarray}}
\def \eea{\end{eqnarray}}
\def \nn{\nonumber}
\def \half{{1\over 2}}

\def \e{{\epsilon}}

\def \L{{\Lambda}}
\def \a{{\alpha}}
\def \t{{\theta}}
\def \b{{\beta}}

\def \d{{\delta}}
\def \w{{\omega}}

\def \e{{\epsilon}}

\def \G{{\Gamma}}

\def \av#1{{\langle#1\rangle}}

\def \beas{\begin{eqnarray*}}
\def \eeas{\end{eqnarray*}}

\def \half{{\frac{1}{2}}}

\newcounter{indice}

\def \bn{\begin{enumerate}}
\def \en{\end{enumerate}}

\def \bb{\begin{thebibliography}{99}}
\def \eb{\end{thebibliography}}

\def \cotgh{{\rm cotgh}}

\begin{document}

\title{Dynamics and universality in noise driven dissipative systems}
\author{Emanuele G. Dalla Torre}
\affiliation{Department of Condensed Matter Physics, Weizmann Institute of Science,
Rehovot, 76100, Israel}
\affiliation{Department of Physics, Harvard University, Cambridge MA
02138}
\author{Eugene Demler}
\affiliation{Department of Physics, Harvard University, Cambridge MA
02138}
\author{Thierry Giamarchi}
\affiliation{DPMC-MaNEP, University of Geneva, 24 Quai Ernest-Ansermet, 1211 Geneva,
Switzerland.}
\author{Ehud Altman}
\affiliation{Department of Condensed Matter Physics, Weizmann Institute of Science,
Rehovot, 76100, Israel}

\date{\today}

\begin{abstract}
We investigate the dynamical properties of low dimensional systems, driven by external noise sources. Specifically we consider a resistively shunted Josephson junction and a one dimensional quantum liquid in a commensurate lattice potential, subject to $1/f$ noise. In absence of nonlinear coupling, we have shown previously that these systems establish a non-equilibrium critical steady state [Nature Phys. 6, 806 (2010)]. Here we use this state as the basis for a controlled renormalization group analysis using the Keldysh path integral formulation to treat the non linearities: the Josephson coupling and the commensurate lattice. The analysis to first order in the coupling constant indicates transitions between superconducting and localized regimes that are smoothly connected to the respective equilibrium transitions. However at second order, the back action of the mode coupling on the critical state leads to renormalization of dissipation and emergence of an effective temperature. In the Josephson junction the
temperature is parametrically small allowing to observe a universal crossover between the superconducting and insulating regimes. The IV characteristics of the junction displays algebraic behavior controlled by the underlying critical state over a wide range. In the noisy one dimensional liquid the generated dissipation and effective temperature are not small as in the junction. We find a crossover between a quasi-localized regime dominated by dissipation and another dominated by temperature. However since in the thermal regime the thermalization rate is parametrically small,
signatures of the non-equilibrium critical state can be seen in transient dynamics.
\end{abstract}
\pacs{05.70.Ln,37.10.Jk,71.10.Pm,03.75.Kk}
\maketitle
\section{Introduction}

Advances in the fabrication and coherent manipulation of quantum many-body systems, such as mesoscopic devices\cite{Braggio},
superconducting circuits\cite{Clerk2010,Clarke2008,Schoelkopf2008} and ultracold atomic gases\cite{Bloch2008} allow to investigate new regimes of
many-body physics and open the way to interesting technological applications. A recurring issue with these systems is that they are often being driven out of equilibrium, either purposely as part of the preparation scheme, or due to their sensitivity to external noise sources. Such studies provide a strong motivation for understanding the properties of quantum many-body systems under non-equilibrium conditions (see e.g. [\onlinecite{Polkovnikov2010b,Zoller-noneq}]) . Since a full microscopic description of such complex systems is in general hopeless, it is important to identify regimes where the complexity of the dynamics gives rise to robust emergent behavior that is independent on the microscopic details.

In this paper we describe universal quantum phenomena that occur in many-body systems subject to external noise sources that are not in thermal contact with the system.
We have previously identified a class of quantum-critical steady states that arise under these conditions \cite{us-nature}. Specifically, we considered quadratic model systems, such as quantum RC circuits and Luttinger liquids at zero temperature subject to $1/f$ noise. Clearly the correlation and response of the systems are modified by the noise and show manifestly non-equilibrium phenomena. But, as we showed, the scale invariance and hence the algebraic decay of the correlations is not destroyed. Non linear perturbations such
as a Josephson junction in the $RC$ circuit or a commensurate periodic potential in the Luttinger liquid were considered using simple scaling arguments. This analysis showed that the perturbations can be tuned from being relevant to irrelevant by varying the noise power, suggesting a possible non-equilibrium phase transition. The key issue that was not addressed by the simple
scaling arguments is the back-action of the non-linear perturbations on the quadratic elements of the system.

Here we extend our previous analysis and use the critical steady states as a starting point for a perturbative real-time renormalization-group (RG) approach. This method was used in Ref. \onlinesquare{GiamarchiMitra} to describe a quench of a Luttinger liquid in a commensurate periodic potential. In the quench problem \cite{GiamarchiMitra,GiamarchiMitra-long} the mode-coupling induced by the lattice potential led to the generation of finite dissipation and temperature. In the present paper, we generalize the RG scheme and extend it to our case of interest namely the steady state regime of quantum systems subjected to $1/f$ noise. This allows us to study the effects of small perturbations around the ``non-equilibrium quantum critical points'' in a controlled way. The first order expansion agrees with the predictions made in Ref.~\onlinesquare{us-nature} by means of the scaling argument. However, the second order shows the appearance of new physics, which was hidden by the trivial scaling analysis.

We highlight the new effects in the example of a ``zero dimensional'' noisy Josephson junction for which we obtain the full dynamical phase diagram. For such a system temperature and friction are generated by the combination of the out of equilibrium noise and the mode coupling terms. This latter effect being absent in the equilibrium system. We also analyze by the present method the noisy one dimensional system of interacting particles. Quite interestingly in the later case, although the microscopic starting point is quite different from the case of quenched systems, the initial steps of the flow are similar to Ref.~\onlinesquare{GiamarchiMitra,GiamarchiMitra-long}.

The plan of the paper is as follows:
the paper starts in Sec.~\ref{sec:models} with a definition of the models we consider, namely the noisy Josephson junction
and the noisy 1D chain. The next section Sec.~\ref{sec:frame} offers a self-contained description of our general RG framework.
This section can be safely skipped by the reader mostly interested by the results and not the actual method used to obtain them.
Such results are discussed in the next two sections. Sec.~\ref{sec:josephson} discusses the case of the noisy Josephson junction.
In particular the dynamical phase diagram and its physical implications have been regrouped in Sec.~\ref{sec:josephson-dyphase} and \ref{sec:josephson-physics}. In a similar way, the noisy 1D chain is examined in Sec.~\ref{sec:1d}, with the physical results in Sec.~\ref{sec:1d-dyphase}. Sec.~\ref{sec:conclusion} closes the article with a summary of our results and an outline of future research directions.

\section{Models} \label{sec:models}

Below we describe the specific models of the noisy Josephson junction and the Luttinger liquid considered in this paper. Both systems consist of three key elements, (i) A quantum system of interacting particles, described by a many-body Hamiltonian $H_{\rm sys}$. (ii) A deterministic noise source, described by a stochastic time-dependent Hamiltonian $H_{\rm noise}(t)$. (iii) A dissipative quantum bath $H_{\rm bath}$.  In what follows, we will assume that the dissipative bath is initially prepared at equilibrium, and due to its large size, does not significantly deviate from this state throughout the whole experiment. Considering only noise averaged quantities, we then obtain a non-equilibrium steady state (NESS), in which correlation and response functions depend only on the time-difference. Due to the constant flow of energy through the system, from the noise source to the zero temperature bath, these correlations and response function explicitly violate the equilibrium fluctuation-dissipation theorem (FDT) and
are derived in the framework of a non-equilibrium Keldysh action, to be described below.

\subsection{Noisy shunted Josephson junction} \label{sec:models-josephson}

\begin{figure}[t]\centering
\includegraphics[scale=0.8]{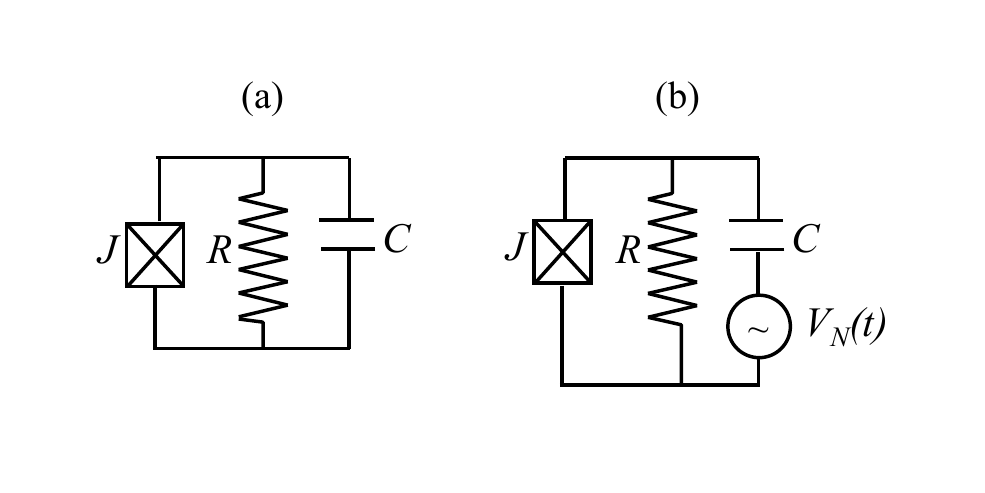}\centering
\caption{Electric circuit of a resistively shunted Josephson Junction (\ref{eq:H}). (a) The junction is at equilibrium. (b) The junction is driven out-of-equilibrium by time-dependent charge noise, modeled by a stochastic voltage source $V_n(t) = C Q(t)$}
\label{fig:sJJcircuit}
\end{figure}

The first system we consider here is the resistively shunted Josephson junction driven by charge noise. This model is defined by the time-dependent Hamiltonian
\be H(t) =  E_J\cos(\theta) + {E_c}\left(n- \frac{Q(t)}{2e}\right)^2 + H_{\rm bath}[n]\label{eq:H}\ee
Here $\theta$ is the phase difference across the junction and $n$ its canonical conjugate (the charge). $E_c=2e^2/C$ and $E_J = \hbar g$ are respectively the charging energy and the Josephson coupling of the junction. $Q(t)$ describes the stochastic time-dependent fluctuations of the off-set charge of the junction (charge noise)\cite{Zimmerli, Schon-dec, Martinis-dec, chargenoise_exp}. $H_{\rm bath}$ models the resistor in terms of an  infinite set of harmonic oscillators \cite{FeynmanVernon,CaldeiraLeggett}. The model can also  be represented by the electric circuit of Fig. \ref{fig:sJJcircuit}, in which the charge noise is generated by a fluctuating voltage source $V_n(t)=C Q(t)$.

Assuming a Gaussian distribution, we average over the noise and obtain an effective action\cite{us-nature} that can be written in the Keldysh form:
\be S = S_{\rm bath} + S_{\rm noise} + S_g\label{eq:SJJ} \nn\ee
\bea
S_{\rm bath}&=& \int\frac{d\omega}{2\pi} \ba{c c}(\theta^*_\w&\hat\theta^*_\w)\\ & \ea G^{-1}_{\rm bath}\left(\ba{c}\theta_\w\\\hat\theta_\w\ea\right)\nn\\
S_{\rm noise}&=& \int\frac{d\omega}{2\pi} \ba{c c}(\theta^*_\w&\hat\theta^*_\w)\\ & \ea G^{-1}_{\rm noise}\left(\ba{c}\theta_\w\\\hat\theta_\w\ea\right)\nn\\
S_g &=& {g}\int dt\sin(\theta)\sin(\hat\theta)
\eea
Here $\theta$ and $\hat\theta$ are the symmetric and antisymmetric combination of the forward ($\theta_+$) and backward ($\theta_-$) paths: $\theta_\pm = \theta\pm\hat\theta$. For an introduction to Keldysh path-integrals see for example Ref.~\onlinesquare{kamenev}. The two fields $\theta$ and $\hat\theta$ are respectively referred
to as the ``classical'' and ``quantum'' fields.

The inverse Green's function of the bath is
\be G^{-1}_{\rm bath}=\left(\ba{c c} 0 & R_QC \w^2 -i\frac{R_Q}{R}\w \\ R_QC \w^2 +i\frac{R_Q}{R}\w &-2i \frac{R_Q}{R}\w\coth\left(\frac{\w}{2T}\right)\ea \right)\ee
Here $R/R_Q=R((2e)^2/h)$ and $T$ are respectively the normalized resistance and $T$ temperature. For 1/f charge noise $\av{Q^*_\w Q_\w} =  (2e)^2 F/|\w|$ and
\be G^{-1}_{\rm noise}=\left(\ba{c c} 0 & 0 \\ 0  & -4\pi i F |\w|\ea \right)\ee
For later convenience we define a scaled noise parameter
\be
{\bar F}\equiv 2\pi (R/R_Q) F\label{eq:F01}
\ee
Note that the noise contributes only to the quantum-quantum part of the action: it introduces only fluctuations in the system, without any dissipation, and therefore explicitly violates the equilibrium fluctuation-dissipation theorem (FDT).

We have shown in Ref.~\onlinesquare{us-nature} that $1/f$ noise preserves the  scale invariance of the quadratic model. As a result physical correlations decay as power-laws in time. Of particular interest are the correlation ${\mathcal{K}}(t)$ and response function ${\mathcal R}(t)$ of the cooper pair creation operator $e^{i\t}$. These are easily calculated within the quadratic theory to be\cite{us-nature}:
\bea
{\mathcal{K}}(t-t') &\equiv& \half\av{e^{i\t_+(t')}\left(e^{i\t_+(t)}+e^{i\t_-(t)}\right)}\approx \left(\frac{RC}{|t-t'|}\right)^{2(1+{\bar F})R/R_Q}\label{eq:mathK}\nn\\
{\mathcal R}(t-t') &\equiv& \frac{i}{2}\av{e^{i\t_+(t')}\left(e^{i\t_+(t)}-e^{i\t_-(t)}\right)}\approx \Theta(t)\sin\left(\pi\frac{R}{R_Q}\right) {\mathcal{K}}(t-t') \label{eq:mathR}
\eea

An important goal of this work is to investigate the effect of the non linear coupling introduced by the Josephson junction on the
steady state of the system. We shall study these effects in Sec.~\ref{sec:josephson} using a real-time RG within the Keldysh framework.
The general RG scheme is described below in Sec.~\ref{sec:frame}.

\subsection{Noisy Luttinger liquid}\label{sec:models-1d}

The second example we consider consists of interacting particles in one dimension, coupled to a quantum bath and to a source of time-dependent noise. The many-body Hamiltonian describing the system, the bath and the noise is:
\begin{multline}\label{eq:HLL}
H(t) = \frac{1}{2\pi}\int \frac{dx}{u}  \left[K\left(u \pi\partial_x\theta(x)\right)^2 + \frac{1}K\left(u\partial_x\phi(x)\right)^2 - g(x)\cos(2\phi)\right] 
+ H_{\rm bath}[\phi] -\frac1\pi \int \frac{dx}{u}  f(x,t)\partial_x\phi(x)
\end{multline}
The first two terms are the standard Luttinger liquid Hamiltonian with Luttinger parameter $K$ and sound velocity $u$. The fields $\theta(x)$ and $-(1/\pi)\phi(x)$  are, respectively, the phase of the bosons and their long-wavelength density fluctuations\cite{Haldane,giamarchi,giamarchi-notes,giamarchireview}. They satisfy the canonical commutation relation: $[-\frac1\pi\partial_x \phi(x),\theta(x)]=i\delta(x-x')$. The cosine term describes an external static potential. In particular, $g(x)=g~u~\delta(x)$ models a local impurity\cite{KaneFisher}, while $g(x)=g$ models a periodic potential at commensurate filling\cite{Jaksch}. The last two terms of (\ref{eq:HLL}) describe respectively a dissipative bath and a stochastic time-dependent noise, linearly coupled to the long-wavelength fluctuations of the density.
In Ref.~\onlinesquare{us-nature} we have discussed specific examples, such as one dimensional systems of ultra-cold polar molecules and chains of trapped ions for which this model is expected to be relevant.

After averaging over the noise, we obtain the following real time action:
\be S = \int \frac{u~dq}{2\pi}~\frac{d\omega}{2\pi} \ba{c c}(\phi*&\hat\phi^*)\\ & \ea G^{-1}_{1d}\left(\ba{c}\phi\\\hat\phi\ea\right) + \int dt~dx~g(x)\cos(2\phi)\label{eq:S1d}\ee
Here
\be G^{-1}_{1d}=
\left(\ba{c c} 0 & \frac1{\pi K}(\w^2- u^2 q^2)+i\eta\w \\ \frac1{\pi K}(\w^2-u^2 q^2)-i\eta\w &
-2i\eta\left(|\w|+ {F\over\eta} \frac{q^2}{|\w|} \right) \ea \right)\label{eq:S1d-quadratic}\ee
Here we assumed a noise spectrum $F(q,\w)=\av{f^*(q,w)f(q,w)}=F/|\w|$, implying that the noise couples to smooth density fluctuations but is uncorrelated at long scales. This behavior of the noise spectrum is consistent with recent measurements of 1/f noise on cold trapped ions\cite{MonroeNoise,ChuangNoise}. The parameter $\eta$ describes the coupling strength to the bath, which is assumed to be linear, ohmic, and at zero temperature. (For the physical motivations of these assumptions, see the Supplementary material of Ref.~\onlinesquare{us-nature}.) For later convenience we also define the dimensionless noise strength
\be
{\bar F}={F\over u^2\pi^2\eta}\label{eq:F02}
\ee

The model (\ref{eq:S1d}) is analogous to the zero dimensional case presented above, (\ref{eq:SJJ}), with one important difference. In eq. (\ref{eq:SJJ}) the coupling to the thermal bath $R/R_Q$ was a dimensionless parameter. Here on the other hand the coupling $\eta$ has units of momentum and is therefore relevant in an RG sense. To access the quantum critical point, we restrict ourselves to the case of a weak coupling between the system and the bath, by considering the limit of $\eta\to0$ (while keeping a fixed ${\bar F}$). Our results are therefore valid only for length-scales smaller than $1/\eta$. In fact, this limitation is often less restrictive than
other infrared cutoffs already existing in system, such as the finite size of the sample.

In the absence of an external potential ($g(x)\equiv 0$), the action is quadratic and one can easily compute any physical observable. Consider for instance the crystalline order parameter $O(x) = \cos(2\phi(x))$, which measures the closeness of the system to a perfect crystal. ($O(x)=1$ in a perfect crystal and $O(x)=0$ in a non-interacting gas). At equilibrium, this order parameter decays algebraically signaling a quasi-long range ordered crystal. In the presence of the noise we have previously shown \cite{us-nature} that the crystalline order is still algebraic but with a modified exponent. Specifically, the correlation and response functions of the crystalline order are respectively given by
\bea
{\mathcal{K}}(x-x',t-t') & \equiv& \half\av{e^{i\phi_+(x',t')}\left(e^{i\phi_+(x,t)}+e^{i\phi_-(x,t)}\right)} \approx \left(\frac{a^2}{(x-x')^2-u^2(t- t')^2}\right)^{K(1+{\bar F})}\label{eq:Cexpcritical2-models}\\
{\mathcal R}(x-x',t-t')  &\equiv& \frac{i}{2}\av{e^{i\phi_+(x',t')}\left(e^{i\phi_+(x,t)}-e^{i\phi_-(x,t)}\right)} \approx \sin(\pi K)\Theta( u(t-t')-|x-x'|){\mathcal{K}}(t-t')\label{eq:Rexpcritical2-models}
\eea
Here $a$ the inter-particle distance (i.e. the natural UV cutoff). If ${\bar F}\neq0$, we obtain a scale invariant state which explicitly violates the equilibrium fluctuation-dissipation theorem: a non-equilibrium quantum critical state.

The static potential introduces a non linear mode coupling term and our goal is to understand its effect on the critical steady state described above.
Let us first discuss the case of a local impurity potential, i.e. $g(x)=g~u~\d(x)$. This can be thought of as a modified Kane-Fisher problem\cite{KaneFisher} where the Luttinger liquid lead is subject to local $1/f$ noise. Similarly to the equilibrium case, this problem can be mapped exactly to the Josephson junction model described above.
To explicitly see the mapping, we integrate-out the Luttinger liquid and obtain an effective action for the field $\phi_\w(x=0)$ on the impurity site. The quadratic part of the action is then simply the inverse Green's function of this field calculated in the quadratic theory (\ref{eq:S1d-quadratic})
\bea
G_K(\w,x=0)&=& i\av{{\phi}^\star_\w(0){\phi}_\w(0)}= i \lim_{\eta\to0} \frac{K}{2\pi} \int (u~dq)~\frac{\eta|\w| + \frac{F q^2}{|\w|}}{|u^2 q^2-\w^2-i\eta\w|^2}\\
&=&i \lim_{\eta\to0} \frac{K}{2}\left(1+\frac{F}{\pi^2\eta}\frac{q^2}{\w^2}\right) \int (u~dq)~\delta(u^2 q^2-\w^2) = i \frac{K}{4}\left(1+{\bar F}\right) \frac{1}{|\w|}\\
G_R(\w,x=0)&=&i\av{{\hat\phi}^\star_\w(0)\phi_\w(0)}=i \lim_{\eta\to 0} \frac{K}{2\pi} \int u~ d q~\frac{1}{u^2 q^2-\w^2-i\eta\w}
= \frac{K}{4i\w}
\eea
Inverting the two by two Green's function matrix we obtain an action that is formally equivalent to the  noisy Josephson junction (\ref{eq:SJJ}) with $R/R_Q=K$ and $C=0$. However we note that the fields appearing in the impurity action have a different physical meaning. In Sec.~\ref{sec:impurity} we will show how to adapt the results derived for the Josephson junction in order to obtain the current-voltage characteristics of the Luttinger liquid with a single impurity.

The case of a commensurate periodic potential (i.e. $g(x)=g$ constant) is different. We shall treat this problem using the RG approach in section \ref{sec:1d}. For now we would like to point out the close similarity between the problem of a noisy Luttinger liquid in a commensurate potential and that of the steady state established after a sudden quench of the Luttinger parameter from $K_i$ to $K_f$ in a sine-gordon model. The latter problem was investigated by Mitra and one of us in Refs. \onlinesquare{GiamarchiMitra,GiamarchiMitra-long}. Without the periodic potential the steady state following the quench, as in our case, exhibits algebraic correlations with a modified exponent that depends on the strength of the quench\cite{Cazalilla}. Indeed the correlations and the response in this state
are fully described by a Keldysh action almost identical to (\ref{eq:S1d}) except for the substitution $({\bar F} q^2/|\w|) \to (K_f/K_i - 1)|\w|$.
When building the RG equations around the Gaussian action this
difference turns out to be unimportant in the limit $\eta\to 0$ because of the relativistic dispersion of the Luttinger liquid. Since the flow
will take the system away from this fixed point, some of the physics of these two problems could however correspond to two different fixed points.

\section{RG framework}\label{sec:frame}

In both models described above, the action of the system can be written as a sum of a quadratic and a non-quadratic term. As noted above and in Ref.~\onlinesquare{us-nature}, the quadratic part of the action is scale invariant and gives rise to a critical steady state. This provides the basis for treating the non-quadratic terms using a perturbative renormalization group (RG) method controlled by the proximity to the scale invariant steady state.

The philosophy of the RG scheme is similar to the equilibrium case. As usual we envision measuring the system with probes of decreasing frequency resolution $\Lambda$. We should be able to compute the measured dynamic correlations using an effective Keldysh action $S_\Lambda$ that has $\Lambda$ as its upper cutoff in frequency instead of the larger physical cutoff $\Lambda_0$. The effective action must be modified in such a way that all dynamical correlations at scales $\w\le\Lambda$ are the same as would be computed using the original action. The flow of the coefficients of the different terms in the action with the decreasing frequency scale $\Lambda$ reveals the physical mechanisms that play a dominant role at each scale.

If the bare action was just the scale invariant quadratic action, then the lower energy
effective action $S_\Lambda$ would of course be the same as the original action after a trivial rescaling
of the coordinates and fields. By contrast,
the non-linear term $g$ couples modes of different frequencies.
Therefore elimination of the fast modes will necessitate a change of the effective action to mimic the effect of the lost high frequency modes on the slow modes through the mode coupling term.  As long as the coupling term $g$ is small, this change is concomitantly small and can be computed by perturbation theory in $g$ at each stage as we gradually eliminate
shells of modes at decreasing frequency.

We now wish to make an important technical note on the implementation of the cutoff
within a real time approach. Within the imaginary time equilibrium formalism, $\Lambda$ is often introduced as a sharp cutoff. This has the advantage of making a clear separation between fast and slow modes at each step of the $RG$.
As we will see, this approach cannot be used for time-dependent problems because it violates causality\cite{chamon}. Instead, similar to Refs. \onlinesquare{Nozieres,Knops}, we implement the cutoff $\L$ by convolving the free Green's function in the quadratic part of the action $G_0(r)$ with a smooth cutoff function $\G_\L=\G(\L r)$.
As we will show, a convenient choice of the cutoff function can significantly simplify the  calculations. The challenge is how to perform the integration over fast modes when there is no sharp separation between fast and slow modes in frequency space. This is done using a technique due to Nozieres and Gallet \cite{Nozieres}, which we adapt below to the non-equilibrium Keldysh framework. The derivation takes similar lines as in Refs.~\onlinesquare{GiamarchiMitra,GiamarchiMitra-long}
by Mitra and one of us. Here we generalize the scheme somewhat in order to allow application to both the one-dimensional system and to the Josephson junction. The main results of this section are summarized by Eq. (\ref{eq:scaleJ}) and Eq. (\ref{eq:GRK}). These two equations give the general forms of the the scaling equations to first order and second order in the strength of the cosine potential.

\subsection{RG of a quadratic action}

As a warm up let us start by considering a quadratic action
\be S_0 = \int d^D r_1 ~d^D r_2~ \phi(r_1) G_0^{-1}(r_1-r_2) \phi(r_2)\label{eq:Squad}\ee
Here $r_i=(\vec {x}_i,t_i)$ is a $D=d+1$ dimensional Keldysh coordinate, running on the forward and backward paths. For a quadratic action, the RG approach is just equivalent to rescaling the coordinates by $r_i \to (1+dl)r_i$. The scaling equations of each parameter appearing in $G^{-1}_0$ are determined by their respective engineering dimension. For example, unitless parameters are scale invariant and define the critical fixed points of the RG flow. Energy scales, like a finite temperature, are always relevant in an RG sense and grow along the RG flow.

At each step of the RG procedure, we need to effectively reduce the cutoff $\L$ by a small amount $d\L = dl~\L$. For this purpose, we express the path-integral weight  $e^{i S_0[\phi]}$ as the integral over an auxiliary field $\d\phi$:
\be e^{i \phi G_\L^{-1} \phi} = \int \mathcal{D}(\d\phi)~ e^{i (\phi-\d\phi) (G_\L-d G_\L)^{-1} (\phi-\d\phi) + i ~\d\phi (d  G_\L)^{-1} \d\phi}\label{eq:RG1} \ee
Here $G_\L = G_0 * \G_\L$ and we used a shortened notation in which the integration over the coordinates is substituted by products of matrices. The function $d G_\L$ implements the change of the cutoff and, for an infinitesimal shift $d\L$, it equals to $d\L \partial_\L G_\L = dl \L \partial_\L G_\L$. We now define a rescaled field $\phi' = \phi - \d\phi$ and re-write the RHS of (\ref{eq:RG1}) as $e^{i S[\phi',\d\phi]}$, where
\be S_0[\phi',\d\phi] =  \int d^D r_1 ~d^D r_2 \left[\phi' (G_\L - d G_\L)^{-1} \phi' + \d\phi (d G_\L)^{-1} \d\phi\right] \label{eq:RG2} \ee
The RG procedure consists of integrating-out the ``fast'' degrees of freedom $\d\phi$. If the action is quadratic, the resulting action for $\phi'$ is just equivalent to the original one, up to a small shift of the cutoff, $G_\L-dG_\L = G_{\L-d\L}$. If, on the other hand, the action contains non-quadratic terms, the integration-out of the fast degrees of freedom can have more important effects, which are the focus of our study.

\subsection{Perturbative RG of a non-quadratic action}\label{sec:RGgen}

Consider now an action made by the sum of two terms: a quadratic term $S_0[\phi]$ and a small non-quadratic term $S_g[\phi]$. We assume that the cutoff is implemented in the quadratic part and transform the non-quadratic part using the definition $\phi=\phi'+\d\phi$. Next, we expand the path integral weight $e^{i (S_0 + S_g)}$ in powers of $S_g$, integrate-out the fast degrees of freedom $\d\phi$ and re-exponentiate the resulting terms. This procedure generates a renormalized action for the ``slow'' fields $\phi'$. Finally, we rescale the coordinates by $r \to (1+dl) r$, to match the original cutoff, and derive closed scaling equations for the parameters of $S_0$ and $S_g$.

In what follows we will apply this scheme to a generic homogeneous non-linear coupling of the form
\be S_g[\phi] = g \int d^D r~ F(\phi(r))\label{eq:Sg}\ee
Here $F = F(\phi)$ is some non-quadratic function of the field $\phi$.

\subsubsection{First order in $g$}

The first order expansion of the Keldysh weight $e^{iS} = e^{i S_0(\L) + i S_g}$ gives $(1+S_g)e^{i S_0(\L)}$. Substituting $\phi=\phi'+\d\phi$ and integrating-out $\d\phi$, we obtain
\be 1 + i \av{S_g}_{\d\phi} = 1 + i~g \int d^D r~ \int \mathcal{D}(\d\phi) F(\phi' + \d\phi) e^{i (\d\phi) (d G_\L)^{-1} (\d\phi)}\ee
For simplicity, we now assume $F(\phi)$ to be analytic and expand it in power-series of $\phi$:
\bea
1 + i\av{S_g}_{\d\phi} &=&  1 + i ~g \int d^D r~ \int \mathcal{D}(\d\phi) \sum_n c_{n} (\phi' + \d\phi)^{n} e^{i (\d\phi) (d G_\L)^{-1} (\d\phi)} \\
&=& 1 +i S_g[\phi'] + i~g \int d^D r  \sum_n \frac{n(n-1)}2 ~ c_{n} (\phi')^{n-2}\int \mathcal{D}(\d\phi) (\d \phi)^2 e^{i (\d\phi) (d G_\L)^{-1} (\d\phi)} \\
&=& 1 +i S_g[\phi'] + \frac{g}2 dG_\L(0) ~\int d^D r~  \partial^2_{\phi} F(\phi') \eea
Here we expanded the binomial $(\phi'+\d\phi)^n$, and we kept only the two lowest order terms in $d G_\L \sim d\L$.

Finally, re-exponentiate the resulting term and rescale the coordinates by $r\to (1+dl)r$.  We obtain that the renormalized action of $\phi'$ differs from the original one by
\be
{d S^{(1)}} = D~dl~S_g[\phi'] + \frac{g}2~dl~\L {\partial \over \partial\L}\av{\phi^2}_\L\int d^D r~ \partial^2_{\phi} F(\phi')\label{eq:first} \ee
The first contribution of (\ref{eq:first}) corresponds to the normal scaling of the coupling, dictated by its engineering dimensions. The second ``anomalous'' contribution can lead to the generation of new couplings in the action, as well as to an anomalous scaling of the existing one.

As expected, if $F(\phi)$ is a quadratic function of $\phi$, $\partial_{\phi^2} F$ is independent of the fields and the anomalous contributions in (\ref{eq:first}) vanishes (due to cancellation of forward and backward paths). If, on the other hand, $F(\phi)$ is a periodic function of $\phi$, the anomalous contribution can have the same functional dependence as the normal one. In the case of a cosine term, for instance, $F(\phi)=\cos(\phi) \Rightarrow \partial^2_{\phi}F = - \cos(\phi)$ and
\be {d S^{(1)}\over dl} = D~ S_g[\phi'] - \L~\half {\partial\over \partial\L}\av{\phi^2}_\L  S_g[\phi'] = \left[D - {\L\over 2}
{\partial\over \partial\L} G_\L(0) \right] S_g[\phi']\label{eq:Sg'}\ee
The resulting scaling equation is
\be \frac{d g}{d l} = g\left[D - \frac{\L}2 \frac{\partial}{\partial\L}\av{\phi^2}_\L\right] \label{eq:scaleJ}\ee
Where $D=d+1$ ($D=1$ in the Josephson junction problem and $D=2$ for the Luttinger liquid). In general, $G_\L(0) = \av{\phi^2}_\L$ is an increasing function of $\L$. As a consequence, the second term in the RHS of (\ref{eq:scaleJ}) tends to counterbalance the first one, eventually making it irrelevant. This is the origin of both quantum phase transitions to be described in the next two sections.

\subsubsection{Second order in $g$}

The first order scaling equations do not include a back action of the interaction term on the quadratic terms. To account for such back action we go on to compute the flow equations to second order in the perturbation g:
\be e^{iS_{0,\L}[\phi] + i S_g[\phi]} \approx (1+i S_g[\phi] - \half S_g^2[\phi]) e^{i S_{0,\L}[\phi]}\ee
As before, we express the field $\phi$ as $\phi'+\d\phi$ and average over the ``fast'' degrees of freedom, $\d\phi$. Next, we re-exponentiate together the first and second order, by using
\be 1+i \av{S_g}_{\d\phi} -\half \av{S^2_g}_{\d\phi} \approx \exp\left[i \av{S_g}_{\d\phi}-\half{S^2_g}+\half\av{S_g}_{\d\phi}^2\right]\label{eq:2ndorder} \ee
Here the average denotes integration over the fast degrees of freedom $\d\phi$, $\av{...}=\av{...}_{\d\phi}=\int d(\d\phi)...e^{i S_{0\L}[\phi',\d\phi]}$. Hence the second order correction to the effective action in terms of $\phi'$ is \be d S^{(2)}[\phi'] = \frac{i}{2}\av{S^2_g}_{\d\phi}-\frac{i}2\av{S_g}_{\d\phi}^2 \ee
As usual, following the integration out of the high-frequency modes we restore the cutoff to its original value by rescaling the coordinates $r\to(1+dl)r$, and then rename the fields, by $\phi' \to \phi$.

Consider for instance the specific case of a homogenous cosine coupling
\bea S_g[\phi] &=& g \int d^D r~ [\cos(\phi_+(r,t))-\cos(\phi_-(r,t))] \\
&=& g \sum_{\e_1,\e_2=\pm}\int d^D r \e_1 \e_2 e^{i\e_1\phi(x,t)+i\e_2\hat\phi(x,t)}\eea
Squaring this action, we obtain
\be S^2_g =\sum_{\e_i}\e_1\e_2\e_3\e_4 \int d^D r_1 \int d^D r_2~  e^{i\Phi_{1234}} \ee
Here again $r$ stands for the $D=d+1$ space time corrdinate and we have defined $\Phi_{1234} = e_1{\phi}_1+\e_2 \hat{\phi}_1+\e_3{\phi}_2+\e_4\hat{\phi}_2$, where $\phi_i = \phi(r_i)=\phi(x_i,t_i)$.

The integration over the fast degrees of freedom and rescaling of the coordinates then gives
\be \av{S^2_g}_{d\phi} = \equiv \frac14{g^2}\sum_{\e_i}\e_1\e_2\e_3\e_4 (1+dl)^{2D} \int d^D r_1 \int d^D r_2 e^{i\Phi_{1234}} \left(1 - dl~\L \frac{\partial\av{\Phi_{1234}^2}}{\partial \L}\right) \label{eq:Sg''}\ee
Using (\ref{eq:2ndorder}) we then find:
\bea
{d S^{(2)}\over d l} &=&~\frac{i}8\L {g^2}\prod_{i=1}^4\sum_{\e_i=\pm 1}\e_1\e_2\e_3\e_4 \int d r_1 \int d r_2 e^{i\Phi_{1234}}\newline {\partial \over \partial \L }\left[\av{\Phi_{1234}^2}_0 - 2\av{{\phi}^2}_0 \right]\label{eq:dS2b}
\eea
This correction to the action includes a renormalization to the quadratic terms of the action that we can find by expanding the exponential factor in small fluctuations around its reference expectation value in the quadratic theory\cite{giamarchi, Nozieres}. Let us define $\d\Phi^2=\Phi^2-\av{\Phi^2}_0$ and expand
\bea
\cos{\Phi} &=& \sum_{n=0}^{\infty} \frac{(-1)^n}{(2n)!}(\av{{\Phi}^2}_0+\d{\Phi}^2)^{n}\nn\\
&=&\sum_{n=0}^{\infty} \frac{(-1)^n}{(2n)!}\left(\av{{\Phi}^2}_0^{n} +n \av{{\Phi}^2}_0^{n-1}\d\Phi^2\right)+O(\d\Phi^4)\nn\\
&=& e^{-\half\av{{\Phi}^2}_0}\left(1 + {\Phi}^2 - \av{{\Phi}^2}_0\right)+O(\d\Phi^4)\label{eq:eiphi}
\eea
The second term in the brackets renormalizes the quadratic action. The other two terms are constant contributions that can be dropped (In fact these terms even cancel-out due to the sum over the $\e_i$s).
We can now substitute the definition of $\Phi_{1234}$ and plug back into Eq. (\ref{eq:dS2b}). In doing so we note that the contribution of terms with $\e_1=\e_3$ vanishes. This is because the exponential $e^{-\half\av{{\Phi}^2}_0}$ includes in this case the factor $\av{({\phi}_1+{\phi}_2)^2}_0\to\infty$. This leads to a restricted sum only over $\e_1,\e_2$ and $\e_3$, which gives
\bw\bea
{d S^{(2)}\over dl} &=& ~~\L~\frac{i}4{g^2}\int dr_1~dr_2~\sum_{\e_1,\e_2,\e_4}\e_2\e_4\left[\e_1({\phi}_1-{\phi}_2)+\e_2\hat{\phi}_1+\e_4\hat{\phi}_2\right]^2\nn\\&&
\times e^{-\half\av{({\phi}_1-{\phi}_2)^2}_0-\half\e_1\e_2\av{\hat{\phi}_1{\phi}_2}}_0 {\partial\over \partial\L} \left[\e_1\e_3\av{{\phi}_1{\phi}_2}_0 + 2\e_1\e_2\av{\hat{\phi}_1{\phi}_2}_0\right]
\eea
We now explicitly perform the summation over $\e_i$ indexes and obtain
\bea
&&{d S^{(2)}\over dl} = -\frac{1}4 {g^2}\int_{-\infty}^\infty d^DR \int d^D r \hat{\phi}(R+{r\over 2})\hat{\phi}(R-{r\over 2}) \L\left[\cos(G_{R,\L}(r)) {\partial G_{K,\L}(r)\over \partial\L}+i\sin(G_{R,\L}(r)) {\partial G_{R,\L}(r)\over \partial\L} \right]\nn\\
&&+\left({\phi}(R+{r \over 2})-{\phi}(R-{r\over2} )\right)\hat{\phi}(R-{r\over 2})\L\left[\sin(G_{R,\L}(r)){\partial G_{K,\L}(r)\over \partial\L}-i\cos(G_{R,\L}(r)) {\partial G_{R,\L}(r)\over \partial\L}\right] e^{i \left(G_{K,\L}(r)- G_{K,\L}(0)\right)}.\label{eq:dS2d}
\eea
where we defined $R=(r_1+r_2)/2$ and $r=r_1-r_2$. $G_{K,\L} = i \av{\phi(t)\phi(0)}$ and $G_{R,\L}= i \av{\phi(t)\hat\phi(0)}$ are the Keldysh and retarded components of the Green's function's in the quadratic action with a cutoff $\L$.

Eq. (\ref{eq:dS2d}) amounts to a renormalization of the quadratic action (\ref{eq:SJJ}). The first row gives the RG transformation of the Keldish component while the second row determines the renormalization of the retarded and advanced components. These second order contributions can be rewritten as
\bea {d G^{-1}_K(r)\over d l} &=& \frac{i}4 ~g^2 \L\left({\partial\over \partial\L} + {\partial \av{\phi^2}\over \partial\L}\right) \cos(G_{R,\L}(r)) e^{i\left(G_{K,\L}(r)-G_{K,\L}(0)\right)}\label{eq:GKbis} \\
{d G^{-1}_R(r)\over dl} &=& \frac{i}4 g^2\L\int dr' \left(\delta(r-r')-\delta(r)\right) \left({\partial\over \partial\L} + {\partial \av{\phi^2}\over \partial\L}\right) \sin(G_{R,\L}(r')) e^{i \left(G_{K,\L}(r)-G_{K,\L}(0)\right)}\label{eq:GRbis} \eea
Here we used $G_K(0)=i\av{\phi^2}$. After a Fourier transform they assume the more compact form
\be {d\over d l}\left(\ba{c c}0 & G_R^{-1}(\w) \\ G_A^{-1}(\w) & G_K^{-1}(\w)\ea\right) = - \frac{\L}{2} g^2~ \left({\partial\over \partial\L} +{\partial \av{\phi^2}\over \partial\L}\right) \left(\ba{c c} 0 & {\mathcal R}(\w) - {\mathcal R}(0)\\{\mathcal R}^*(\w) - {\mathcal R}^*(0) &  i {\mathcal{K}}(\w)\ea\right)\label{eq:GRK} \ee
Here $\w$ is the Fourier conjugate to $r$ and ${\mathcal K}(r)$ and ${\mathcal R}(r)$ are respectively the correlation and response function of the operator $e^{i\phi}$ in the quadratic action $S_0$, already defined in (\ref{eq:mathK}) and (\ref{eq:mathR})

\section{The noisy Josephson junction}\label{sec:josephson}

We now analyze the model of the noisy Josephson junction defined by Eq. (\ref{eq:SJJ}). We first derive the RG equations of the Josephson junction valid up to second order in the Josephson coupling $g$. We shall then analyze the flow in order to obtain a dynamical phase diagram and predict the expected current-voltage characteristics of the junction in the different regimes. We will show that there are wide regimes in which the junction exhibits manifestly non-equilibrium but nevertheless universal behavior that is controlled by the non-equilibrium critical state of the noisy RC circuit.

\subsection{Derivation of the RG equations}\label{sec:josephson-RG}
Here we give a detailed derivation of the RG equations valid to second order in the Josephson coupling constant $g$. Readers that are uninterested in the technical details can skip directly to the summary of the RG flow given by Eqs. (\ref{eq:scaleg}) to (\ref{eq:scaleF}).

Consider first the quadratic action of the $RC$ circuit without the Josephson coupling (${g}=0$). In this case, the scaling dimension of each term of the action is just given by their engineering dimensions. The dimensionless resistance $R/R_Q$ and the dimensionless noise strength $F$ are scale invariant and define a {\it non-equilibrium} critical fixed point.  However, if the resistor is kept at a non-vanishing temperature, then the temperature scales like energy and is always relevant in the RG sense, driving the system away from the above non-equilibrium critical state. The  capacitance $C$ is irrelevant in an RG sense and determines the ultraviolet cutoff of the theory: $\L = 1/RC$. In what follows, we will neglect this term, and substitute it by a generic cutoff function $\G_\L$.

\subsubsection{RG flow to first order in $g$}
We now treat the effects of a weak Josephson coupling (${g}\ll\L$), through the perturbative RG approach presented in the previous section. The first order scaling equation of the coupling $g$ is given by Eq. (\ref{eq:scaleJ}) derived in section \ref{sec:RGgen} for a generic model.

We focus at first on the zero temperature case in which the pure $RC$ circuit would be critical. In this case, the anomalous contribution to the scaling dimension can be easily computed from:
\bea
{\partial\over\partial \Lambda}\av{\phi^2(t)}&=& 2~\frac{R}{R_Q}(1+{\bar F}){\partial\over\partial\Lambda} \int_{-\infty}^\infty d\w \frac{1}{|\w|} \G \left(\frac{|\w|}{\L}\right)\nn\\
&=&- 2~\frac{R}{R_Q}(1+{\bar F}) \frac{1}{\L}\int_0^\infty d\w ~\frac{\partial}{\partial \w} \G\left(\frac{\w}{\L}\right)
= \frac{2}{\L}~\frac{R}{R_Q}(1+{\bar F}).\label{eq:generalcutoff}
\eea
Here ${\bar F}$ is defined in (\ref{eq:F01}). In the last identity we used the defining properties of the cutoff function, namely that $\G(x)\to 1$ for $x\to 0$ and $\G(x)\to 0$ as $x\to \infty$. As expected the final result is independent of the precise form of the cutoff function.  Substituting (\ref{eq:generalcutoff}) into (\ref{eq:scaleJ}), we obtain the scaling equation of the Josephson coupling, valid to first order in $g$:
\be \frac{d{g}}{dl} = g \left(1-\frac{R}{R_Q}\left(1+{\bar F} \right)\right) \ee
This result agrees with the simple scaling arguments given in Ref.~\onlinesquare{us-nature}.

Let us now reintroduce the finite temperature of the bath. As we noted already, a finite temperature is a relevant perturbation, leading already at tree level to the scaling equation $dT/dl=T$. Besides this, the most important effect of a finite temperature is on the first order
renormalization of the coupling $g$. If we take $T > 0$, the integral in equation (\ref{eq:generalcutoff}) is substituted by:
\be K_1 ={\partial\over \partial\L} \int_0^\infty d\w~\frac{1}{\w}\left(\Gamma_\L(\w)+\Gamma_\L(-\w)\right){\rm cotgh}\left(\frac{\w}{2T}\right)\ee
If, for instance, we choose a cutoff of the type $\Gamma_L(\w)=\exp(-|\w|/\L)$, we obtain:
\be K_1 = 2 \frac{1}{\L^2}\int_0^\infty d\w e^{-\w/\L}{\rm cotgh}\left(\frac{\w}{2T}\right) = 2 \frac{T}{\L}\left[\psi\left(\frac{T}{2\L}\right) - \psi\left(\frac{T+\L}{2\L}\right)\right]-2\label{eq:K1}\ee
Here $\psi(x)$ is the Digamma function. For small temperatures $T\ll \L$, $K_1\approx 2$, and we retrieve the zero temperature limit (\ref{eq:generalcutoff}). For high temperatures, on the other hand, the integral is proportional to $K_1 \approx \frac{T}{\L}$. The precise form of the scaling function is clearly cutoff-dependent, while its asymptotic behavior is general and gives $K_1 = 2 \left(1+\frac{T}{2\L}\right)$.


\subsubsection{RG flow to second order in g}

As shown in Sec.\ref{sec:RGgen}, the second order contributions renormalize the quadratic part of the action, according to (\ref{eq:GRK}). To extract explicit scaling equations of the different couplings, we proceed with taking the low frequency expansion of (\ref{eq:GRK}). The zero frequency contribution of the Keldysh part is
\be {d G^{-1}_K(\w=0)\over dl} = \int dt~{d G^{-1}_K(t)\over dl} \equiv -i \frac{g^2}{\L} I_0(R/R_Q,{\bar F})\label{eq:GKter}\ee
Here the dimensionless factor $I_0$ equals to
\bea
I_0 &=& \L^2 \int dt~\left[{\partial\over \partial\L} + \left( {\partial G_{K,\L}(0)\over \partial\L}\right)\right]~{\mathcal K}(t)\\
&=&  \L^2 \int dt~\left[\frac{t}{\L}\partial_t + \left( {\partial G_{K,\L}(0)\over \partial\L}\right)\right]~{\mathcal K}(t)\label{eq:I0-interm}\\
&=& \left(1-D_g\right)\L \int dt ~{\mathcal K}(t)\label{eq:I0-general}\eea
where $D_g=d \ln g/d l=1-R/R_Q(1+{\bar F})$ is the scaling dimension of the junction $S_g$
in the scale invariant steady state of the quadratic action with $1/f$ noise and the resistor at zero temperature.
In writing (\ref{eq:I0-interm}) we used the fact that under these conditions the only scale is provided by the cutoff $\L$ and
therefore ${\mathcal K}(t)$ is a scaling function of $\L t$. We then integrated by parts and used (\ref{eq:scaleJ}) to obtain the
result (\ref{eq:I0-general}).
In appendix \ref{app:josephson} we compute $I_0$ for a specific choice of the cutoff (\ref{eq:I0}). Here we just note  that $I_0$ vanishes when the noise parameter ${\bar F}$ tends to zero and is positive for any $\bar F \neq 0$.

The zero frequency component of the Keldysh action $G_K^{-1}(\w=0)$ can be associated with an effective temperature that is generated in the RG flow. This connection is established by comparing to the low frequency expansion of the Keldysh component of a quadratic action at equilibrium  $\eta\w ~\coth(\w/2T) = \eta T + O(\w^2)$, where $\eta$ is the dissipation (In our case $\eta=R_Q/R$).
Hence we have the mapping $G_K^{-1}(\w=0) \leftrightarrow \eta T$.
The next term in the low frequency expansion of (\ref{eq:GKbis}) is proportional to $\w^2$. Comparing the coefficient of this term with the low frequency expansion of the equilibrium form will not in general give the same effective temperature as inferred from comparison of the leading term.  This reflects the fact that the system is not truly at equilibrium. However in the RG sense the $\w^2$ and higher order terms are irrelevant. Therefore the asymptotic low frequency behavior is that of an effective equilibrium system.

Note that the corresponding correction was computed using the bare action. As the flow
proceeds one should in principle compute this correction with the full renormalized action, and thus in particular
with a finite temperature. This is however very complicated, in particular given the frequency dependence of the generated
Keldysh term. We will thus stick to the correction computed at $T=0$, keeping in mind that at some scale the flow will
be cut by the generated finite temperature. Note that if the system is really in equilibrium (${\bar F}=0$) then $I_0=0$ as it must be. This ensures that a zero temperature system will not generate a finite temperature in the course of renormalization.

We turn to consider the renormalization of the off-diagonal components of the Keldysh action given by Eq. (\ref{eq:GRbis}). The lowest order term in the low-frequency expansion is proportional to $i\w$:
\be {d G^{(-1)}_R\over dl} =  - i\w \int dt~t~{d G^{-1}_R(t)\over dl}  \equiv -i\w \frac{g^2}{\L} I_1({\bar F},R/R_Q)\label{eq:GRter}\ee
The coefficient $I_1$ is given by
\bea
I_1 &=& \L^3 \int dt~t~\left[{\partial\over \partial\L} + \left( {\partial C_\L(0)\over \partial\L}\right)\right]~{\mathcal R}(t)\\
&=& - D_g {dl}\L^2 \int dt~t ~{\mathcal K}(t)\label{eq:I1-general}
\eea
This integral is calculated explicitly in appendix \ref{app:josephson}.
By comparing (\ref{eq:GRter}) with the original action (\ref{eq:SJJ}) we find that $I_1$ is associated with renormalization of the dissipation. Specifically we can rewrite Eq. (\ref{eq:GRter}) as $d\eta/dl=I_1 g^2/\L^2$, where $\eta\equiv R_Q/R$. We note that at equilibrium (${\bar F}=0$) $I_1$ vanishes, which implies that then there is no renormalization of the dissipation. This is in agreement with Refs. \onlinesquare{bulgadaev,FisherZwerger,KaneFisher}.

Renormalization of the dissipative part in the off diagonal component of the action implies a similar renormalization of the external noise term. Recall that the $|\w|$ term in the Keldysh component was $-2i\eta|\w|-4\pi i F|\w|$ and as we saw above it is not renormalized to second order in $g$. We can rewrite this term as $-2i(\eta+d\eta)|\w|-4\pi i (F-d\eta/(2\pi))|\w|$.
In this way the renormalization of the resistor in the off diagonal term is complemented by the required renormalization in the diagonal term so as to maintain the same form as the original action (\ref{eq:SJJ}).
The price is having a renormalization of the noise term, which can be written as:
\be
{d{\bar F}\over dl}=-\frac{g^2}{\L^2}\frac{R}{R_Q}(1+{\bar F}) I_1
\ee
The constant $I_1$ is always positive, meaning that the second order contribution acts to reduce the non-equilibrium noise. Accordingly, the renormalization of the dissipation is accompanied by a corresponding negative renormalization of the temperature.

\subsubsection{RG equations and physical consequences}

The above results can be summarized by the following scaling equations:
\bea
\frac{d {g}}{d l} &=& {g}\left[1- \frac{R}{R_Q}\left(1+\frac{T}{2\L_0}+ {\bar F}\right)\right]\label{eq:scaleg}\\
\frac{d T}{d l} &=& T +{R\over R_Q} \frac{{g}^2}{\L_0} I_0-T{R\over R_Q}\frac{{g}^2}{\L_0^2}I_1\label{eq:scaleT} \\
\frac{d R}{d l} &=& -\frac{R^2}{R_Q}\frac{{g}^2}{\L_0^2}I_1\label{eq:scaleR}\\
\frac{d {\bar F}}{d l} &=& (\a-1){\bar F} - \frac{R}{R_Q}(1+{\bar F})\frac{{g}^2}{\L_0^2}I_1 \label{eq:scaleF}\\
\eea
Here $\Lambda_0\approx 1/(RC)$ is the bare cutoff of the junction. The unitless coefficients $I_0$ and $I_1$ are defined in (\ref{eq:GKter}) and (\ref{eq:GRter}). As shown in appendix \ref{app:josephson}, it is possible to obtain analytic expressions for both coefficients, by an appropriate choice of the cutoff.

Let us now review the different features of the equations (\ref{eq:scaleg}-\ref{eq:scaleF}) which describe the RG flow of the noisy shunted Josephson junction (\ref{eq:SJJ}) up to second order in $g/\L$. The flow of the Josephson coupling term given by Eq. (\ref{eq:scaleg}) is completely analogous to the equilibrium case. The first term in the brackets reflects the engineering dimension of the Josephson coupling, which has units of energy while the remaining terms imply an anomalous scaling dimension. At equilibrium and zero temperature ($T=0$, ${\bar F}=0$) this equation reduces to the well known result of Refs.~\onlinesquare{Schmidt,Chakravarty} with the anomalous contribution to the scaling dimension being $-R/R_Q$. Both the
temperature and the noise {\it decrease} the anomalous dimension, making the
Josephson coupling less relevant. At this level the picture is identical to that
given in Ref.~\onlinesquare{us-nature}, using a simple scaling argument.

New physics enters in the renormalization of the circuit at second order in $g/\L$. In particular
Eq.~\ref{eq:scaleT} implies that a finite effective temperature of order $g^2/\L$ is generated in the course of renormalization even if the temperature of the resistor is zero at the outset. Physically this can be understood as follows. In absence of a mode coupling term the energy imparted on the system by the noise is distributed exactly as dictated by the noise spectrum. For example the spectrum of fluctuations induced by $1/f$ noise is proportional to $|\w|$, very different from a thermal distribution. The effect of mode coupling is to redistribute the mode occupation and in particular it generates a constant contribution (i.e. frequency independent) to the distribution of fluctuations in the zero frequency limit. The spectral weight of fluctuations at low frequencies and hence the effective temperature, can be estimated using Fermi's golden rule. First it is proportional to $g^2/\L$ which is related to the matrix elements for transitions from occupied modes at high energies to the low energies. Second, the generated effective temperature is proportional to $I_0$, as defined in (\ref{eq:GKter}), which is related to the non-equilibrium population of high frequency modes. As required $I_0$ tends to zero at equilibrium ${\bar F}=0$. The value of the effective temperature is set by the ratio between the fluctuations in the zero frequency limit and the dissipation $R_Q/R$.

There is also a reciprocal process due to the mode coupling term, that acts to renormalize the dissipation as described by Eq. (\ref{eq:scaleR}). It is worthwhile to write this equation directly in terms of the dissipation parameter $\eta=R_Q/R$ as
\be
{d\eta\over d l}={g^2\over\L^2}I_1
\ee
Physically, when there is a non equilibrium population of high frequency modes they can absorb energy from the low frequency modes by changing their population and thereby serve as an effective dissipative bath. The added dissipation is proportional to the coupling between the modes $(g/\L)^2$ and to the coefficient $I_1$,  which is a measure of the deviation from equilibrium ($I_1=0$ for ${\bar F}=0$). Note that, as shown in the Appendix~\ref{app:explicitRG}, $I_1$ diverges for $(1+{\bar F})R/R_Q<1/2$. In this regime, the above low frequency expansion becomes invalid. We expect that, in this regime, a non-analytic term $\sim \w^{\mu}$ with $\mu<1$ is generated, which corresponds to sub-ohmic dissipation. The physical implications of this transition deserve further study, which is postponed to future work.

The last equation (\ref{eq:scaleF}) describes the scaling of the external noise.
$1/f$ noise is marginal at the level of the quadratic action, but it becomes weakly irrelevant at second order in the mode coupling term. This is consistent with the system flowing toward an effective equilibrium state with a finite effective temperature.

\subsection{RG equations at strong coupling: Duality transformation}

So far we have treated the case of a weak Josephson coupling in a perturbative RG scheme to second order in the coupling $g$.
The limit of a strong junction can be addressed by means of a well known duality transformation\cite{Schmidt,FisherZwerger},
which is carried out in detail within the Keldysh framework in Appendix~\ref{app:duality}.
In this case the phase across the junction is essentially locked except in rare occasions when a quantum fluctuation induces a phase slip,
which can be formally described as a tunneling event of a dual-particle across a weak dual junction.
The current of dual particles is the rate of phase slips that gives the voltage drop on the original Josephson junction.

We can thus regard the dual junction as embedded in a complete dual circuit that is identical in structure to the original circuit
with the transformations.
\bea
g&\to& g_{dual}\nn\\
R&\to& R^{-1}\nn\\
F&\to& (R/R_Q)^2 F\nn\\
T&\to& T
\label{eq:duality}
\eea
The prefactor of the dual cosine, $g_{\rm dual}$, can be  evaluated using W.K.B. approximation as\cite{Schmidt} $g_{\rm dual} \approx \L \exp\left(-\sqrt{g/\L}\right)$ . In particular, if $g\gg \L$, then $g_{dual}$ is small, facilitating the perturbative RG approach in $g_{dual}$. The RG equations are exactly the same as (\ref{eq:scaleg})-(\ref{eq:scaleF}) above except with the
replacements dictated by the duality transformation (\ref{eq:duality}).

\subsection{Dynamical phase diagram}\label{sec:josephson-dyphase}


 The RG flow equations derived above for the weak and strong coupling limits can serve to identify the different steady state regimes
 of the system. We will show in this section that the entire parameter space
is divided into three distinct regimes, which can be best described as (i) non-equilibrium superconducting, (ii) non-equilibrium insulating and
(iii) thermal metal.  We shall identify the characteristic scales that define the different regimes and derive analytic expressions for the
crossover lines which separate them. 

We start the analysis of the dynamical regimes from the weak coupling limit,
that is when the (bare) Josephson coupling is much smaller than the cutoff $\Lambda_0=1/(RC)$.
A rough separation into a superconducting and an insulating regime can be made using the first order flow of the coupling $g$ determined by Eq. (\ref{eq:scaleg}) with $T=0$ 
\be \frac{d {g}}{d l} = {g}\left[1- \frac{R}{R_Q}\left(1+ {\bar F}\right)\right]\label{eq:scalegT0}\ee
At this order we would identify the insulating phase as the region $R/R_Q<(1+{\bar F})^{-1}$, where the Josephson coupling is irrelevant, and the
superconducting phase as the complementary region where $g$ flows to strong coupling.

The picture becomes somewhat more involved when
we consider the second order correction to the flow. Most importantly, Eq. (\ref{eq:scaleT}) shows that in presence of external noise, the Josephson coupling acts to generate an effective temperature. There is always a scale $\Lambda_T$ below which the temperature becomes the dominant term and
where the junction will exhibit metallic behavior. But that scale may be so small that $g(\L)$ meanwhile changes by orders of magnitude from its bare value. If $g(\Lambda)$ flows to strong coupling, i.e. exceeds the cutoff $\L_0$, at a scale $\Lambda_S$ larger than the thermal scale $\Lambda_T$ then there is a wide range of frequency scales over which the junction exhibits universal superconducting behavior.
Similarly in the insulating side of (\ref{eq:scalegT0}), if $g(\Lambda)$ decreases by an order of magnitude at a scale $\Lambda_I$ larger than the thermal scale then there is a wide frequency range over which the Junction exhibits insulating behavior.

Let us now estimate the scales $\Lambda_I,\Lambda_S$ and $\L_T$. To this end we first solve equation
(\ref{eq:scalegT0}) to obtain $g(l) = g_0 \exp\left((1-{\tilde R})l\right)$, where we have defined ${\tilde R}\equiv (1+{\bar F})R/R_Q$.
When $g$ is irrelevant $(\tR>1)$ we get the typical (logarithmic) scale $l_I=(\tR-1)^{-1}$ and therefore $\L_I=\L_0\exp(-(\tR-1)^{-1})$.
If on the other hand $g$ is relevant $(\tR<1$) it reaches the cutoff at the scale $\Lambda_S=\Lambda_0(g_0/\L_0)^{1/(1-\tR)}$.
Now to find the thermal scale we substitute $g(l)$
 into the flow equation (\ref{eq:scaleT})
 for the temperature. We then solve for $T(l)$ with the initial condition
of zero temperature $T(0)=0$ to obtain
\be T(l) = \frac{g_0^2}{\L_0}\frac{I_0 R/R_Q}{2\tR-1}\left(1-e^{(1-2\tR)l}\right)e^l \approx  \frac{g_0^2}{\L_0}\frac{I_0 R/R_Q}{2\tR-1} e^l.\label{eq:solT}\ee
The last equality is for large $l$ and $\tR>1/2$ which is the regime of interest in any case.
From here we immediately obtain the thermal scale
\be
\L_T= I_0\left({R/R_Q\over 2\tR-1}\right){g_0^2\over\L_0}\equiv T_{\rm eff}
\ee

The crossover between the superconducting regime and the
thermal regime in the system parameter space $R/R_Q$, ${\bar F}$ and $g_0/\L_0$ is defined by the surface
$\L_T=\Lambda_S$ which is given by
\be
{g_0\over\L_0}=\left({I_0 R/R_Q\over 2(1+{\bar F})R/R_Q-1}\right)^{1-(1+{\bar F})R/R_Q\over 2(1+{\bar F})R/R_Q-1}\label{eq:cross1}
\ee
At larger values of $g_0$ the junction flows to strong coupling and displays superconducting behavior over a wide
frequency range.

We can find the crossover between the thermal regime and the insulating regime as the surface
on which the thermal scale $\L_T$ equals $\L_I$. This crossover is given by
\be \frac{g_0^2}{\L_0^2} = \frac{2(1+{\bar F})R/R_Q-1}{I_0 R/R_Q}e^{\left({1-(1+{\bar F})R/R_Q}\right)^{-1}}\label{eq:cross2}\ee
For smaller values of $g_0$ the junction displays an insulating behavior controlled by the weak coupling critical point. For larger $g_0$, the effective temperature smears out the insulating behavior.

\begin{figure}[t]\centering
\includegraphics[scale=0.6]{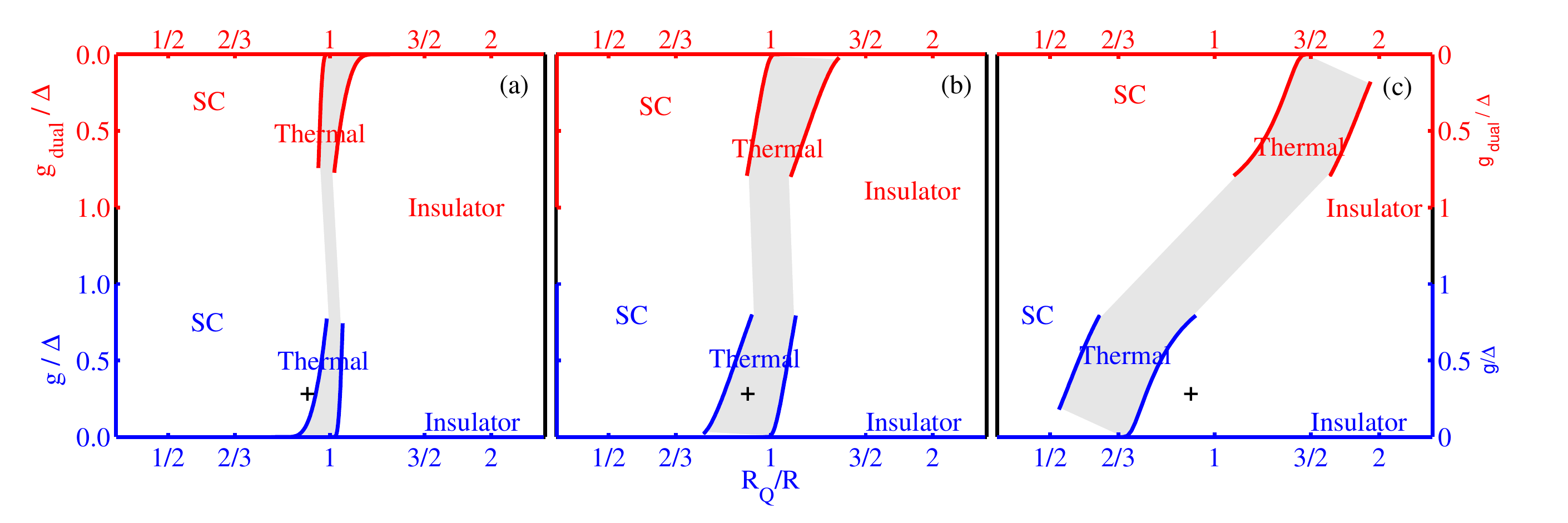}\centering
\caption{Phase diagram of the noisy Josephson junction, as function of
the bare resistance $R/R_Q$ and the bare Josephson coupling $g_0/\L_0$,
for different values of the noise: (a) ${\bar F}=0.01$, $(b) {\bar F} = 0.1$, (c) ${\bar F} = 2.0$. The blue lines
(lower half of each subfigure) are obtained from the weak coupling
analysis and correspond to equations (\ref{eq:cross1}) and
(\ref{eq:cross2}). The red lines (upper half of each subfigure)
are obtained through the duality transformation. The crosses ($+$) indicate the points used for the numerical solution of the RG equations shown in Fig.\ref{fig:RGflow}.}
\label{fig:phdia}
\end{figure}

If the bare Josephson coupling is large with respect to the cutoff scale $\L_0$, one can apply the duality transformation described above and obtain an effective description of the junction in terms of a small phase-slip fugacity. In the dual picture, the roles of the insulating and superconducting regimes are exchanged. Therefore the dual version of eq. (\ref{eq:cross1}) describes the crossover between the insulating and thermal regime, and that of eq. (\ref{eq:cross2}) the crossover between the thermal and superconducting regimes.

The phase diagram derived in this way is shown in Fig. \ref{fig:phdia} for both the weak coupling and strong coupling regimes for different values of the noise power. We see that the noise has two main effects on the phase diagram: first it shifts the transition from superconductor to insulator away from the universal value $R_c=R_Q$. The shift is in opposite directions in the weak and strong coupling regimes.  Second, increasing noise leads to a growing crossover region which exhibits effective thermal equilibrium behavior between the two states. In the next section we shall derive the universal transport characteristic expected in each of the three regimes.

\begin{figure}[t]\centering
\includegraphics[scale=0.6]{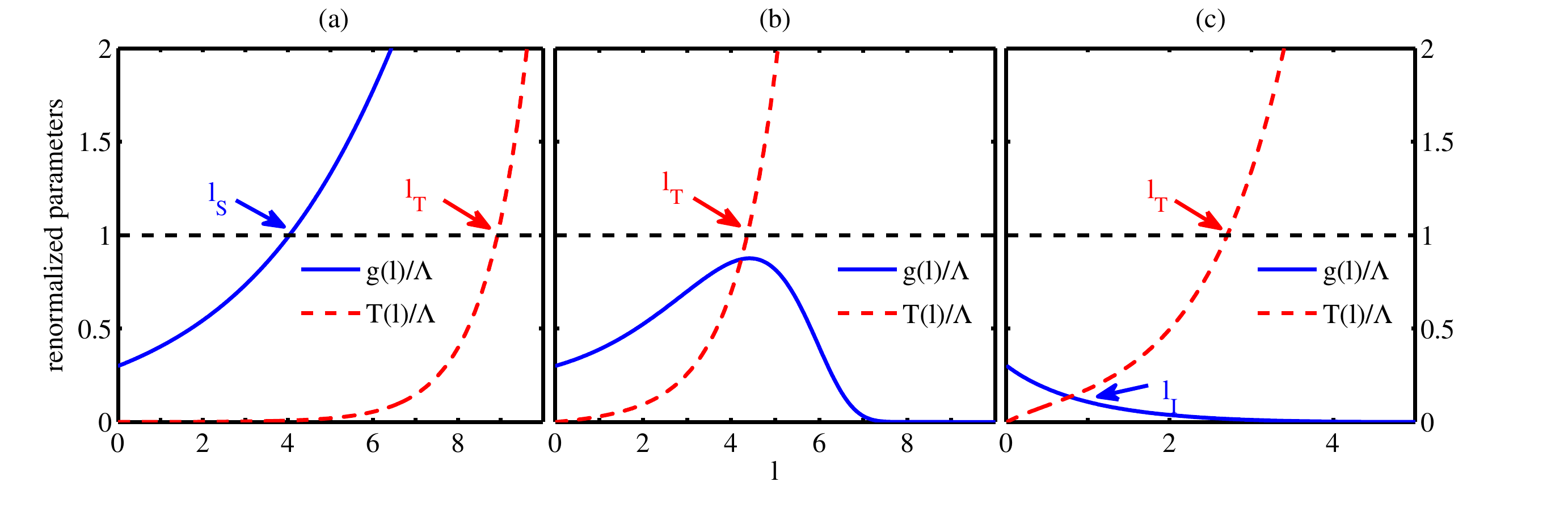}\centering
\caption{Example for RG flows of the Josephson coupling $g(l)$ and of the effective temperature $T(l)$ in the different
dynamical regimes. We used the parameters $g_0/\Lambda_0=0.3$, $R/R_Q=0.7$, $T_0=0$ and different noise parameters { (a)} ${\bar F}=0.001$, happens to be in the Superconducting regime. The Josephson coupling reaches the cutoff scale $\Lambda_0$ and runs off to strong coupling before any appreciable temperature is generated. { (b)} ${\bar F}=0.1$ is in the thermal metal regime. The Josephson coupling may grow initially, but the effective temperature is first to reach the cutoff scale. At lower frequency scales the Josephson coupling is strongly suppressed by the effect of temperature. { (c)} ${\bar F}=2.0$ is in the insulating regime.The Josephson coupling is irrelevant and it has a chance to decrease significantly from its bare value before the temperature reaches the cutoff scale.}\label{fig:RGflow}
\end{figure}

To confirm the phase diagram obtained from the approximate solution of the RG equations, we solve the equations (\ref{eq:scaleg}-\ref{eq:scaleF}) numerically for various values of the bare parameters. Figure (\ref{fig:RGflow})  shows an example of the RG flow for the following set of parameters $g_0/\L_0=0.1$, $R_0/R_Q=0.7$, $T_0=0$ and ${\bar F}=0.01,0.1,1.0$. These points where chosen to demonstrate a noise-driven transition and are located respectively in the superconducting, thermal and insulating regimes. We see that in the superconducting regime (a) the Josephson coupling grows rapidly and reaches the cutoff scale ($\L$) before the temperature does. In the thermal regime (b) the temperature grows more rapidly than the Josephson coupling, preventing the latter from reaching the cutoff scale. Finally, in the insulating regime (c), the Josephson coupling decreases significantly, before the temperature becomes dominant.

\subsection{Current voltage characteristics of the junction}\label{sec:josephson-physics}

The different dynamical regimes determined from the RG flow are characterized by distinct transport
properties. Here we derive the I-V characteristics of the junction. We find a simple universal
behavior in the different regimes that nevertheless betrays the non-equilibrium nature of the system.

\subsubsection{Insulating regime}

The insulating state occurs in the regime where the Josephson coupling is irrelevant. We therefore expect the transport
properties to be well described by a perturbative analysis in the weak coupling limit.
The standard expression for the current through a voltage-biased shunted Josephson junction, to second order in the Josephson coupling $g$
is
\bea
I&=&I_n + I_s = \frac{V}R + g^2{\text Im} {\mathcal R}(\w=eV)
\eea
Here $I_n$ and $I_s$ are respectively the normal current and the super-current through the junction. ${\mathcal R}(\w)$ is the response function of the operator $e^{i\t(t)}$ in the steady state of the $RC$ circuit unperturbed by the Josephson junction: it corresponds to the Fourier transform of (\ref{eq:mathR}). This perturbative calculation is sometimes referred to as the $P(E)$ approach in the context of Josephson junctions \cite{IngoldNazarov}. For voltages significantly higher than the thermal scale $\Lambda_T$, we can use the zero-temperature result (\ref{eq:GRter}) to obtain
\be I_{s}(V) = \frac{g^2}{\L^2}  I_1 V \label{eq:IVpert}\ee
Here $I_1=I_1(R/R_Q,F)$ is a unitless constant, whose explicit expression is given in Eq. (\ref{eq:I1})

\begin{figure}[t]\centering
\includegraphics[scale=0.65]{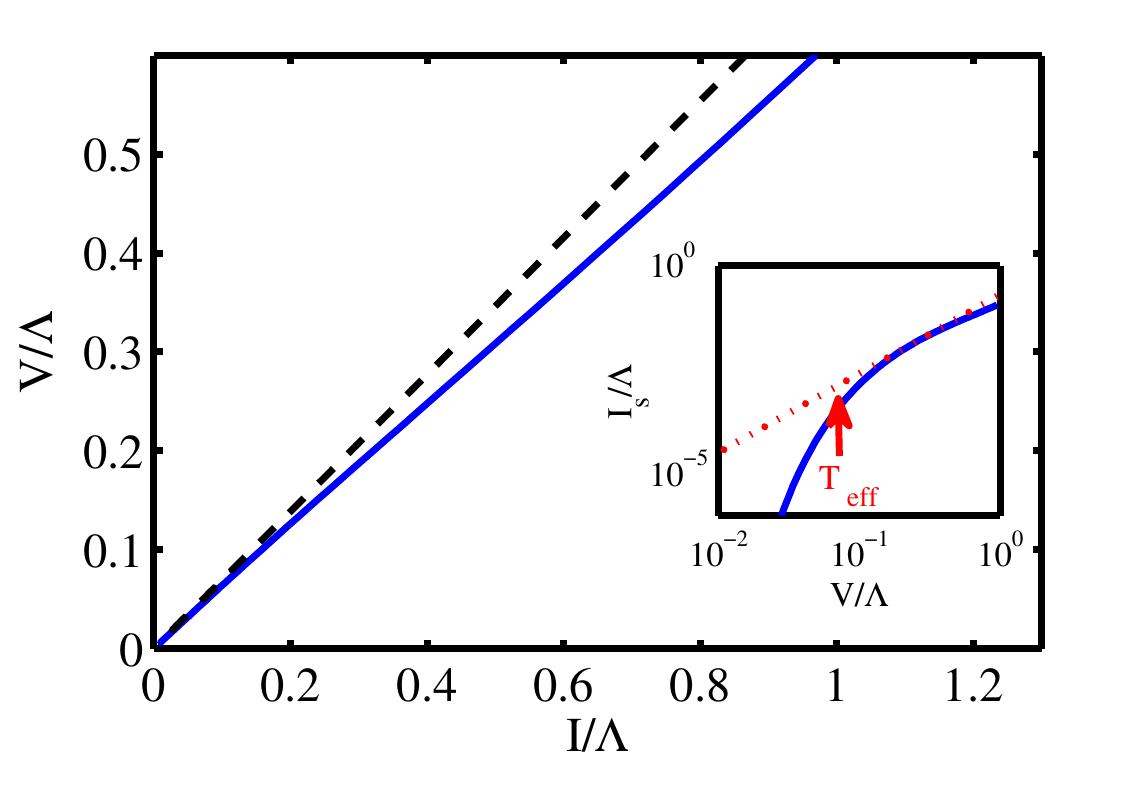}
\caption{I-V characteristic of the Josephson junction in the insulating regime of the phase diagram Fig. \ref{fig:phdia}, calculated from the RG flow using Eqs. (\ref{eq:IVins-RG}). The bare parameters of the junction are the same as the one chosen for Fig.\ref{fig:RGflow}(c): $g/\L=0.3$, $R/R_Q=0.7$, and ${\bar F}=2.0$. If the junction was a perfect insulator all the current would flow through the resistor and the curve would be the dotted line $V=R I$. In reality the junction carries a non-vanishing current $I_s$ leading to a power law deviation from the linear curve. INSET: double logarithmic plot of the supercurrent $I_s$. The dotted line corresponds to an ideal power law with exponent $R(l_T)/R_Q(1+\bar{F}(l_T))$. The actual $I_s$ follows the power-law down to the effective temperature $T_{\rm eff}=\Lambda e^{-l_T}$} \label{fig:IVcurve1bis}
\end{figure}

The above perturbative result can be improved by utilizing the RG approach. Instead of using perturbation theory on the bare model we apply it to the renormalized model at the scale $\Lambda=V$ at which point the RG flow must be terminated. If we neglect the temperature, this gives
exactly the expression (\ref{eq:IVpert}), only with $\Lambda \to V$ and renormalized values of parameters
\bea
I_{s}(V)&=&  I_1(l_V) \frac{g^2(l_V)}{V} \\
I_{n}(V) &=& \frac{V}{R(l_V)}.
\label{eq:IVins-RG}
\eea
Here $I_1(l) = I_1(R(l)/R_Q,{\bar F}(l))$ and $\l_V=\log(\L/V)$. We see from this that the $I-V$ curve is a power-law only at
voltages higher than the thermal scale, at smaller voltages, the Josephson coupling is suppressed by thermal fluctuations and the contribution of cooper pair tunneling to the total current becomes negligible.
In addition, the RG flow shows that the algebraic decay is actually controlled by the renormalized resistance $R=R(l=\log(V))$ and noise ${\bar F} = {\bar F}(l=\log(V))$, rather than by their bare values. This correction is especially important if the bare Josephson coupling is large $g_0 \gg \L$ and the insulating regime is reached only as a consequence of a long RG flow. If, on the other hand, the bare Josephson coupling is small $g_0 \ll \L$, both the corrections to $R$ and to ${\bar F}$ are negligible, and the perturbative result coincides with the RG result.


\subsubsection{Superconducting regime}

\begin{figure}[t]\centering
\includegraphics[scale=0.75]{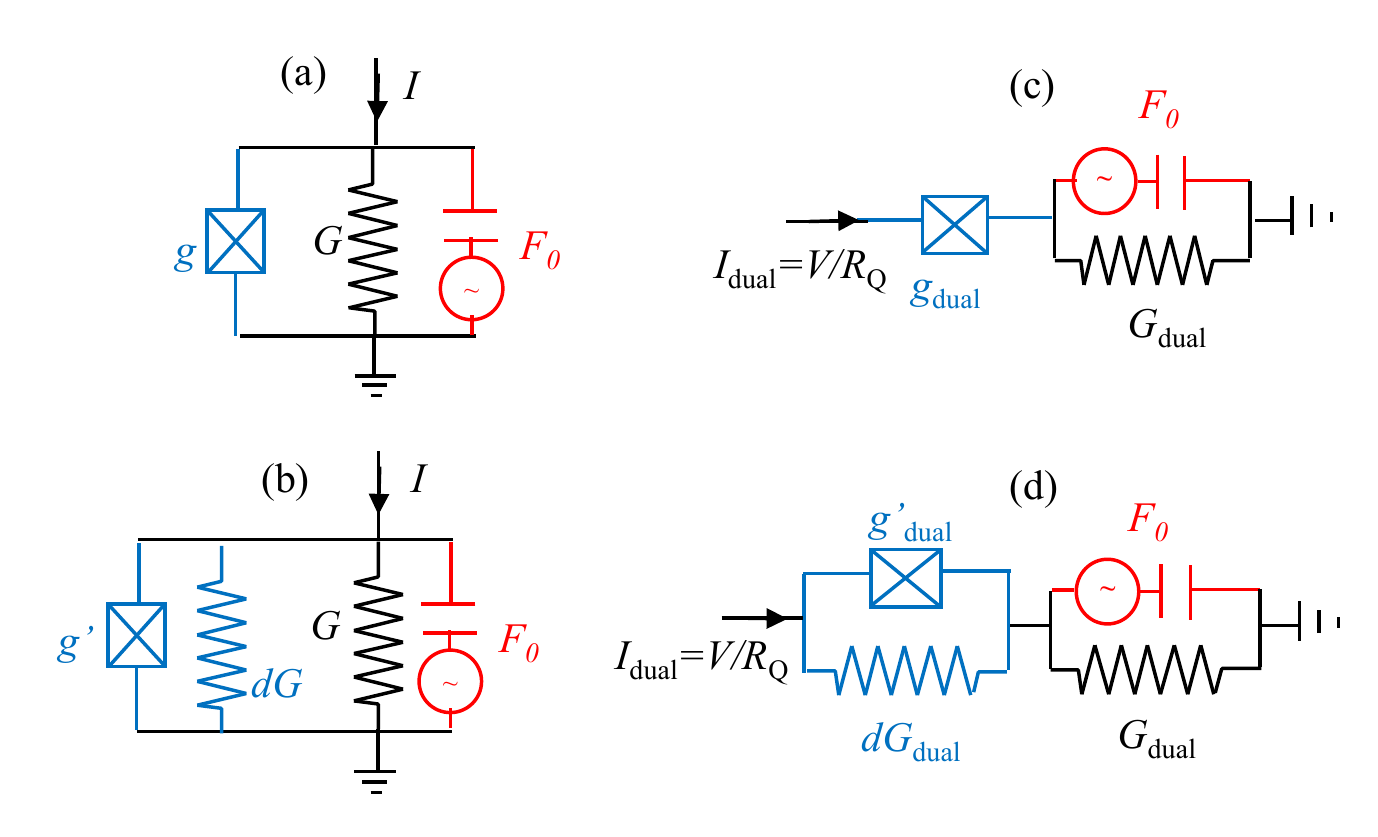}
\caption{Sketch of the effective circuits used for calculating I-V characteristics. (a) is just the bare physical circuit. Panel (b) shows the renormalized circuit including the effective conductance $\d G$ that is generated in the course of renormalization. Panel (c) shows the bare circuit corresponding to the dual theory and (d) is the renormalized dual circuit. Note that the dual current $I_{dual}$ can be thought of as the current of flux quanta trasversing perpendicular to the junction and the resistor in panel (a).}\label{fig:SJJdual}
\end{figure}

In the superconducting regime the Josephson coupling runs off to strong coupling well before reaching the generated thermal scale. Therefore a perturbative analysis
of the superconducting regime can be made from the dual standpoint in which the phase slip fugacity is the small parameter.

We start from a direct perturbative analysis applied to the bare dual circuit. To understand the connectivity of the dual circuit, note that a voltage bias of the original circuit corresponds to a constant current $I_{\rm dual}=V/e$ of flux (dual particles) which flows perpendicular to the original circuit, first across the junction and then across the resistor. The current through the original circuit corresponds, on the other hand, to the sum of the dual voltages on the junction and on the resistor. Hence, in the dual picture we have a resistor $R_{\rm dual}=R_Q^2/R$ connected in series with a Josephson junction $g_{dual}\ll \Lambda_0$. This is sketched in Fig. \ref{fig:SJJdual}.  (See also appendix \ref{app:duality} for a detailed derivation of this result).

We can now apply the weak coupling result (\ref{eq:IVpert}) to obtain $I_{\rm dual}$ versus $V_{s,\rm dual}$ on the dual Josephson element alone. Then using $I_{\rm dual}=V$ and $V_{s,\rm dual}=I$ and substituting the dual junction and noise parameters we can translate the result to physical voltage versus cooper pair current on the original Josephson junction.
\be
V= I_1\left(\frac{R_Q}{R},{\bar F}\right) \frac{g^2_{\rm dual}}{\L} \left(\frac{R_Q I_s}{\L}\right)^{2\frac{R_Q}{R}\left(1+{\bar F}\right)-1}\label{eq:IVpert2}
\ee

Before applying the RG approach to compute I-V relations we should understand how the connectivity of the dual circuit changes in the course of renormalization. Under the renormalization we generate an effective dissipation, which can be regarded as added conductance in the circuit. The (dual) current that went through the bare Josephson junction, now splits between the junction and the
conductance added in the renormalization process. This implies that the added effective dissipation is connected in parallel to the dual Josephson junction and both are connected in series to the resistor $R_{dual}(0)$ of the bare circuit as depicted in Fig. \ref{fig:SJJdual} (b).

Now to compute the $I-V$ relation we take the voltage on the dual circuit as the cutoff to the RG
flow $\Lambda_*=V_{dual}$. The total (dual) current is then given by:
\be
I_{dual}=\d G_{dual}(l_*) V_{dual}+ I_1(l_*) {g_{dual}(l_*)^2\over V_{dual}}
\ee
Where $\d G(l)$ is the renormalized conductance in parallel to the JJ at the scale $l$.
The last relation can be immediately translated to one between the physical voltage and current
\be
V=\d G_{dual}(l_*) I+  I_1(l_*) {g_{dual}(l_*)^2\over R_Q I}
\label{eq:Vdual}
\ee
Note that $\d G_{dual}$ is proportional to $I_1 g_{dual}^2/\Lambda$, hence both contributions to the voltage on the Josephson junction are of order $g_{dual}^2$ as might have been expected. The first term in (\ref{eq:Vdual}) constitutes a small linear resistance, that accompanies the emergence of effective temperature and dominates the transport at very low currents ($I/2e < T_{eff}$). In the other limit, when the current is much higher than the effective temperature, the second term in (\ref{eq:Vdual}) dominates and leads to essentially the same result as bare perturbation theory (\ref{eq:IVpert2}). 

\begin{figure}[t]\centering
\includegraphics[scale=0.65]{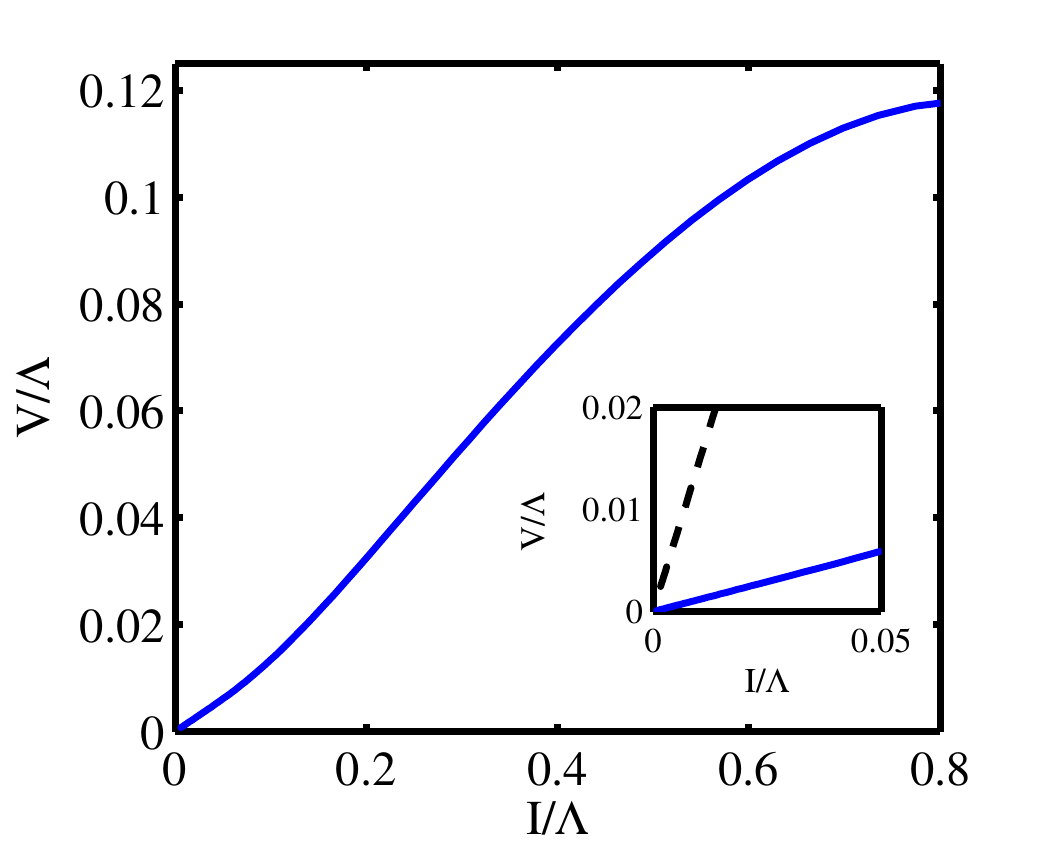}
\caption{I-V characteristic of a noisy Josephson junction in the superconducting regime for  $g_{\rm dual}/\L_0=0.3$, $R_Q/R=0.7$, $F=2.0$. In the regime $T_{eff}\ll I/2e \ll \L_{g,\rm dual}$ we see the algebraic behavior, predicted by perturbation theory in the strong coupling limit (\ref{eq:IVpert2}). In the regime $I/2e\ll T_{\rm eff}$, highlighted in the inset, the junction follows Ohm's law with an effective resistance $\d R = \d G_{dual}$, perturbatively smaller than the bare resistance $R$ (dashed line)}\label{fig:IVcurve2bis}
\end{figure}

\subsection{Impurity in a Luttinger liquid}
\label{sec:impurity}

As we have seen in Sec.~ \ref{sec:models-1d}, the non-equilibrium action of a local impurity in a one dimensional system driven by 1/f noise can be mapped exactly to the action considered here (\ref{eq:SJJ}). This mapping allows us to use the results from the previous section to determine the non-equilibrium phase diagram and IV curves of the one dimensional case. The same mapping allows us to compute the current-voltage characteristics as well. In the 1d model, it is possible to induce a constant flow of particles  by the transformation $\phi(x,t) \to \phi(x,t) + \pi I t$. (Recall\cite{giamarchi-notes} that the displacement field $\phi(x_0,t)$ jumps by $\pi$ each time that a particles crosses the point $x=x_0$). The flow of particles (passing through both the one dimensional system and the impurity) is therefore equivalent to a voltage over the shunted Josephson junction (acting on both the junction and the transistor).  To introduce a constant voltage in the 1d system, we need to add the following term to the Hamiltonian
\be V\int_{-\infty}^{x_0}dx~\rho(x_0) = \frac{V}{2\pi} 2\phi(x_0)\label{eq:Voltage}\ee
Here $\rho(x)\approx (-1/\phi)\partial_x\phi(x)$ is the density of particles at position $x$. Eq.~\ref{eq:Voltage} shows that the voltage of the 1d system (given by the sum of the voltage over the 1d lattice and the voltage over the single impurity) corresponds to the current through the noisy Josephson circuit (given by the sum of the current through the resistor and the super-current through the junction). Using this mapping, we find that current-voltage (IV) characteristic of the noisy 1d system with a local impurity is precisely identical to the voltage-current (VI) curve of a noisy shunted Josephson junction that we computed in Sec.~\ref{sec:josephson-physics}.

\section{One dimensional system}
\label{sec:1d}

We turn to discuss the model of interacting particles in one dimension in a commensurate periodic potential subject to $1/f$ noise. The model was introduced in Ref.~\onlinesquare{us-nature} and reviewed in Sec.~\ref{sec:models-1d}. We have pointed out that in absence of the periodic potential the system exhibits algebraic crystalline correlations and response functions that may be very different from the equilibrium ones. The power law decay of the correlations depends on the dimensionless ratio between the noise strength and the coupling to the bath, ${\bar F}=\pi^{-2}F/(u\eta)$. Recall that we are considering the limit where both $\eta$ and $F$ tend to zero but the ratio between them ${\bar F}$ may be arbitrary.

Having found a scale invariant state in the quadratic model, we should ask how the non-linear coupling of (\ref{eq:S1d}) behaves under the same scaling transformation. In Ref.~\onlinesquare{us-nature} we read off the anomalous scaling dimension of the cosine term from the power-law decay exponent of the crystalline correlations (\ref{eq:Cexpcritical2-models}), which gives $dg/dl=g \left(2-K\left(1+{\bar F}\right)\right)$.
This result suggests a possible noise-tuned  phase transition in which the periodic
potential $g$ changes from being irrelevant to being relevant. In equilibrium (${\bar F}=0$) such a  quantum pinning transition indeed occurs, and as can be seen from the scaling of $g$, it occurs at the universal value of $K=2$.

The purpose of this article  is to go beyond the simple scaling argument and account for the back action of the periodic potential on the steady state correlations. This will be done using the general RG framework laid out in section \ref{sec:frame}.
As mentioned already, there is a formal similarity between this problem and that of the steady state following a sudden change of the Luttinger parameter in the Sine-Gordon model. The RG equations for the latter problem were derived in Refs.~\onlinesquare{GiamarchiMitra,GiamarchiMitra-long}. We shall briefly rederive these equations here for completeness and analyze the dynamical regimes they give rise to in the noisy Luttinger liquid.

\subsection{RG equations}

The general framework that has been laid out in Sec. \ref{sec:RGgen} can be directly applied to the one-dimensional system. First we derive from the general formula (\ref{eq:scaleJ}) an explicit scaling equation for the coupling $g$ valid to first order in the coupling. Then we
use the general formula (\ref{eq:GRK}) for the scaling to second order in $g$. We carry out an expansion of this result in small frequency $\w$ and wavevector $q$ in order
to recast the functional equation (\ref{eq:GRK}) as scaling equations for the set of parameters in the action (\ref{eq:S1d}).
Readers uninterested in the technical details of the derivation can skip directly to equations (\ref{eq:scaleLL0})- (\ref{eq:scaleLL4}) that summarize the result.

One technical note we should make before proceeding with the, is that for the derivation in section \ref{sec:RGgen} we assumed a cosine coupling of the form $\cos(\phi(x,t))$, while we are now interested in the coupling $\cos(2\phi(x,t))$ appearing in the action (\ref{eq:S1d}). We should therefore make the  substitution $\phi\to 2\phi$ in the formulas appearing in Sec. \ref{sec:RGgen} when considering this problem.

To find the first order flow of $g$ explicitly, we need to compute the variation of the fluctuation $\av{\phi^2}$ within the quadratic theory changing cutoff scale. Neglecting the effect of the dissipation $\eta$ we get
\bea \frac{\partial}{\partial\L}\av{\phi^2} & =& \frac{\partial}{\partial\L}\lim_{\eta\to 0}  \int (u~dq)~d\w~ e^{-u|q|/\L} \frac{\eta\w \cotgh\left(\frac{\b\w}2\right)+ q^2 \frac{F}{\w}}{\left|\frac1{\pi K}(\w^2-u^2 q^2)+i\eta\w\right|^2}\\
& =& K \int (u~dq)~ d\w~u|q|~e^{-u|q|/\L}~\left(\cotgh\left(\frac{\b\w}2\right)+\frac{F}{\pi^{2}\frac{q^2}{\w^2}\eta}\right) \delta(\w^2-(uq)^2)~\\
&\approx& K\left[1+\frac{T}{\L} +{\bar F}\right]\eea
Here we used an exponential cutoff function $\Gamma_\L(q,\w)=\exp(-u|q|/\L)$ and ${\bar F}$ is defined in (\ref{eq:F02}). Substituting this result in (\ref{eq:scaleJ}) leads to the scaling equation given in (\ref{eq:scaleLL0}) below.

We now turn to derive the flow equations to second order in $g$ through an expansion of the functional equation (\ref{eq:GRK}) in small $q$ and $\w$.
A contribution with both $q=0$ and $\w=0$ is found only in the Keldysh component $d G_K^{-1}(q,\w)$ and is given by
\bea
\frac{d G_K^{-1}(\w=0)}{dl} &=& i \frac{2g^2}{\L^2}~\left(\L\frac{\partial \av{\phi^2}}{\partial \L} -1\right)  ~\int d(x/u)~\int dt ~ {\mathcal K}(x,t)\\
&=&  i \frac{2g^2}{\L^2}~\left(2-2D_g\right)  ~\int d(x/u)~\int dt ~ {\mathcal K}(x,t) \equiv  -2i\frac{g^2}{\L^2}I_T({\bar F},K)\label{eq:IT-general}\eea
Here $D_g$ is the scaling dimension of the commensurate potential found from the first order flow for vanishing temperature and dissipation, $D_g=2-K(1+{\bar F})$. ${\mathcal K}(x,t)$ is the correlation function of the operator $e^{i2\phi}$ taken in the quadratic action with $T=0$ and $\eta=0$. We used the fact that under these conditions the correlation  is a scaling function of $\L u t$ and $\L x$, ${\mathcal K}={\mathcal K}(\L x/u, \L t)$. As discussed already below Eq. (\ref{eq:I0-general}), this contribution to the Keldysh component can be associated with an effective temperature in the low frequency limit through $G^{-1}_K(0,0)\leftrightarrow \eta T$. The identification is made by comparing to the low frequency expansion of the
Keldysh component at equilibrium $\eta\w ~\coth(\w/2T) = \eta T + O(\w^2)$.

By inversion and time reversal symmetry $d G_K^{-1}(x,t)$ does not contain terms proportional to $q$ or $\w$, and its next non-vanishing terms are proportional to $q^2$ and $\w^2$. These terms are less relevant than the bare quadratic action, and therefore they do not affect the correlation and response functions at long times.

In contrast to the Keldysh component, the renormalization of the off-diagonal component of the action $d G_R(q,\w)$ does not have a zero frequency component. The lowest order term in the gradient expansion of this term in (\ref{eq:GRK}) is
\bea \frac{d G_R^{-1}(\w)}{dl} &\approx& i \w \frac{g^2}{\L^3} ~\left(2\frac{d}{dl}\av{\phi^2} -3\right)~\int dx~\int dt~ t {\mathcal R}(x,t)\\
&=& i \frac{g^2}{\L^2}~\left(1-2D_g \right)  ~\int d(x/u)~\int dt ~ {\mathcal R}(x,t)  \equiv i \w \frac{g^2}{\L^3}I_\eta({\bar F},K) \label{eq:Ieta-general}\eea
This contribution amounts to renormalization of the dissipation as $d\eta/dl = i g^2/\L^4 I_\eta({\bar F},K)$.

The next order in the gradient expansion of $d G_R^{-1}$ gives terms proportional to $q^2$ and $\w^2$. These terms renormalize the bare term $(u K)\w^2 - (u/K)q^2)$. The independent renormalization of $K$ and $u$ are then given by
\be\frac{dK^{-1}}{dl}= -2 dl \frac{g^2}{\L^4}D_g~\int d(x/u)~\int dt~ ((x/u)^2+t^2) {\mathcal R}(x,t)\equiv dl~ \frac{g^2}{u^2\L^4} I_K({\bar F},K)\label{eq:IK-general}  \ee
\be\frac{du}{dl} = -2 dl \frac{g^2}{\L^4} D_g~\int d(x/u)~\int dt~ ((x/u)^2-t^2) {\mathcal R}(x,t)\equiv dl~ \frac{g^2}{u^2\L^4} I_u({\bar F},K)\label{eq:Iu-general} \ee

The above results are summarized in the following RG scaling equations:
\bea
\frac{d g}{d l} &=& g \left[2-K\left(1+\frac{T}{\L_0}+\frac{F}{\pi^2\eta}\right)\right]\label{eq:scaleLL0}\\
\frac{d \eta}{d l} &=& \eta + \frac{g^2}{\L_0^3} I_\eta \label{eq:scaleLL1}\\
\frac{d (\eta {\bar F})}{d l} &=& - \frac{g^2}{\L_0^3} I_\eta(1+{\bar F}) \label{eq:scaleLL1bis}\\
\frac{d (\eta T)}{d l} &=& 2 \eta T + \frac{g^2}{\L_0^2} I_T \label{eq:scaleLL2}\\
\frac{d K^{-1}}{d l} &=& \frac{g^2}{\L_0^4} I_K \label{eq:scaleLL3} \\
\frac{d u}{d l} &=& \frac{g^2}{\L_0^4}I_u\label{eq:scaleLL4}
\eea

The coefficients $I_T$, $I_\eta$, $I_K$, $I_u$ are defined in (\ref{eq:IT-general}-\ref{eq:Iu-general}). Analytic formulas for these coefficients as functions of $K$ and ${\bar F}$ are derived in Appendix~\ref{app:1d}. It should be noted that precisely the same scaling equations were obtained in Ref.~\onlinesquare{GiamarchiMitra-long} in the context of a quench of a one dimensional system in the presence of a commensurate potential. (See also Sec.~\ref{sec:models-1d}).

Let us briefly point to the key properties of the above scaling equations. The first order flow of the coupling $g$
described by Eq. (\ref{eq:scaleLL0}) is very similar to the equilibrium situation but with a negative contribution of the noise through ${\bar F}$ to the anomalous scaling dimension. It is important to note that at this order in the RG the noise, although weakening the correlation functions, is not acting as a temperature, since it preserves the power-law behavior of the correlations.
One might thus expect some change in the dynamics of the particles from delocalized to localized when the noise is tuned to change $g$ from being irrelevant to relevant. At this level, the physics is identical to that described by the simple scaling argument in Ref.~\onlinesquare{us-nature}.

As in the case of the Josephson junction new physics is introduced by effects that are second order in the coupling $g$.
Eq.~\ref{eq:scaleLL1} for example leads to the emergence of finite dissipation $\eta_{eff}\sim I_\eta g^2/\L_0^3$ in the steady state although the bare dissipation ($\eta(0)$) was infinitesimal. The analytic expression given in Eq.~\ref{eq:Ieta} shows that the coefficient $I_\eta$ vanishes identically at equilibrium ${\bar F}=0$, but also that in presence of the noise the effective dissipation diverges for $\tilde{K}\equiv K(1+{\bar F}) <2$, that is exactly where $g$ becomes relevant.
We can interpret this result as emergence of sub-ohmic dissipation that is not captured by the perturbative $RG$ scheme.
In what follows we restrict ourselves to the regime ${\tilde K}>2$, where $\eta_{\rm eff}$ can be made as small as we want by taking a sufficiently weak lattice. The renormalization of the dissipation generates also an effective renormalization of the noise (\ref{eq:scaleLL1bis}), introduced to conserve constant the Keldysh part of the action.

The emergence of dissipation is accompanied by the generation of an effective temperature described by Eq. (\ref{eq:scaleLL2}). We can estimate this temperature from the effective fluctuation-dissipation relation between
the Keldysh and retarded parts of the action
\be
T_{\rm eff}\approx I_T g^2/(\eta_{eff} \L^2)= \L_0 I_T/I_\eta\sim \L_0 ({\tilde K}-2)
\ee
This is a rather suprising result as it implies that the temperature generated in the steady state may be large even for arbitrarily small non-linear coupling $g$ and noise ${\bar F}$. Furthermore, if we track the generated temperature as a function of the flow parameter we see that the effective temperature reaches its large value immediately at the beginning of the
flow. That is, after one infinitesimal RG iteration we have $T(l=0^+)=T_{\rm eff}$. From this point of view it looks like
the non-equilibrium critical state that exists in the purely quadratic model $g=0$ is an extremely singular limit.
We shall discuss below the implications of this result and argue that the critical state of the quadratic model does nevertheless in practice govern the physics of the system with a weak non-linearity.

Note that in the case of the Josephson Junction discussed earlier the effective temperature was parametrically small in both the noise strength and the non-linear coupling. The difference between the two cases stems from the different nature of the dissipation. While the effective dissipation in the one-dimensional system is small, of order $g^2$, in the Josephson junction model even the bare dissipation is of order 1.

The scaling equations (\ref{eq:scaleLL3}) and (\ref{eq:scaleLL4}) for the dimensionless couplings $K$ and $u$ are controlled by the integrals $I_K$ and $I_u$, defined in (\ref{eq:IK}) and (\ref{eq:Iu}). Though we do not have analytical expressions for these coefficients, numerical evaluation of the integrals shows that both are finite for any $\tilde K > 2$. Moreover the factors $I_K$ and $I_u$ do not vanish in the equilibrium case ${\bar F}=0$.
This is expected given that $K$ should also have a non trivial flow near the Kosterlitz-Thouless transition of the Sine-Gordon model at equilibrium (see for example Ref.~\onlinesquare{giamarchi}).

\subsection{Non-equilibrium phase diagram}\label{sec:1d-dyphase}

Analysis of the RG equations (\ref{eq:scaleLL0}-\ref{eq:scaleLL4}) suggests that there are at least
two distinct dynamical regimes in presence of $1/f$ noise (${\bar F}>0$). First, for $\tilde{K}=K(1+{\bar F})<2$
the dissipation $\eta$ diverges through the divergence of the coefficient $I_\eta$. This probably indicates generation of local sub-ohmic dissipation which is beyond the present RG scheme. For ${\tilde K}>2$, an effective temperature $T_{\rm eff}\sim ({\tilde K}-2)\L_0$ is generated immediately at the outset of the flow. We will now consider each regime separately and then move to the crossover region between them.

\subsubsection{Critical - thermal regime}

To understand the nature of the thermal regime, recall that the source of thermalization in the system is the mode-coupling term $g$, which scatters the non-equilibrium population of high-frequency modes towards lower frequencies. The rate by which this scattering occurs is the generated dissipation rate $\eta_{\rm eff}\approx I_\eta g^2/\L$. This rate sets the time scale for the system to achieve the steady state if the non-linearity were turned on at some point in time. But, once the system reaches steady state, its effective temperature $T_{\rm eff}$  depends only on the energy density in the initial critical steady-state and not on $g$.
Thus the singularity of the state at $g=0$ is resolved by the time-span over which the experiment takes place.
In order to see the true steady state with temperature $T\sim ({\tilde K}-2)\L_0$ we would need to
conduct an experiment over a time that diverges as $\tau\sim 1/g^2$ as $g\to 0$.
As a consequence, if we restrict to time scales shorter than $1/\eta_{\rm eff}$, the system is governed by the quadratic part of the action (\ref{eq:S1d}) with negligible $T$. This is a quasi steady state, with scale invariant correlations and response given by (\ref{eq:Cexpcritical2-models}) and (\ref{eq:Rexpcritical2-models}). 

\subsubsection{Local dissipation}

As the effective Luttinger parameter decreases and approaches $\tilde K=2$, the renormalization of the dissipation diverges and the dissipation constant inevitably becomes the first to reach the cutoff. The scale at which this happens is the generated effective dissipation. Clearly at that scale the dissipation is the largest scale in the system and we can neglect the original kinetic term $K^{-1} \w^2$ compared to it. Therefore the effective action we should work with, below the dissipations scale $\L_\eta=\eta_{\rm eff}$ is:
\be S_{\L_\eta}(q,\w)=
\left(\ba{c c} 0 &-{u^2\over K\pi}q^2+ i\L_\eta \w \\-{u^2\over K\pi}q^2-i\L_\eta \w &
-2i\eta_0|\w|+F \frac{q^2}{|\w|} +T(\L_\eta)\ea \right)+ S_g\label{eq:S1d-local}\ee
This describes the thermal regime of a critical theory with dynamical critical exponent $z=2$. It is interesting to note that in this theory the noise becomes a relevant perturbation with scaling dimension $D_F=1$, exactly like the temperature.

\subsubsection{Crossover regime}

We have identified above two dynamical regimes. For $K<2$ the renormalization of the dissipation diverges, which leads to a thermal state with short range spatial correlations already at initial stages of the flow, that is even at frequency scales. On the other hand, for $K\gg2$ there is a wide range of frequency scales in which the dissipation is still small. If the finite time of the experiment does not allow for equilibration to the high effective temperature of the steady state then the physics will be governed by the non equilibrium critical state of the Luttinger liquid.

Of course at sufficiently small frequencies, the generated dissipation being a relevant perturbation will eventually dominate. So, the transition between the two regimes is not perfectly sharp but rather a smooth crossover. Nevertheless we can chart the line of the crossover in the space of ${\tilde K}$ versus the non-linearity $g$ by invoking similar arguments we used above for the Josephson junction. Namely, we should find the regime in ${\tilde K},g$ space, where a critical flow of $g$ down by an order of magnitude can be observed before the dissipation reaches the cutoff scale. where $g$ changes by an order of magnitude.

The typical scale at which $g$ is reduced by an order of magnitude in absence of  dissipation is found by solving the scaling equation for $g$, assuming no renormalization of the dissipation zero temperature and. The solution
\be g(l) = g_0 e^{(2-\tilde K)l}\label{eq:solg2}\ee, gives the frequency scale
$\L_g = \L e^{(2-\tilde K)^{-1}}$.
Now to determine the typical scale where dissipation enters significantly
we solve the scaling equation for $\eta$ (\ref{eq:scaleLL2}) while substituting
the above solution for $g$. This gives
\be \eta(l) = I_\eta\frac{g^2}{\L^3}\left(1-e^{2(2-\tilde{K})l}-1\right)e^{l}\label{eq:soleta2} \ee
\be \L_\eta = \frac{g_0^2}{2\L^3}\frac{I_\eta}{\tilde K -2} \ee
Here we assumed $I_\eta$ to be constant, neglecting the renormalization of the Luttinger parameter and of the sound velocity. Recalling that $I_\eta \propto (\tilde K -2)^{-1}$, we find that as we move away from the transition point $\tilde K =2$, the effective dissipation $\L_\eta$ decays like $(\tilde  K - 2)^{-2}$.

Comparing $\Lambda_\eta$ and $\Lambda_g$, we obtain an analytic expression for the crossover line between the dissipative regime and the critical regime:
\be \frac{g^2}{\L^4} = \frac{2\tilde K -4}{I_\eta} e^{(2-\tilde K)^{-1}}\label{eq:cross3} \ee
For smaller values of $g$, the dissipation scale is smaller than the scale for $g$ in the critical state and it is possible to observe the critical regime. At $\tilde K\to 2$, the transition line goes to zero: any infinitesimal amount of non-linearity is enough to generate an extremely strong dissipation. At large $\tilde K$ the effective dissipation is small, but, as will see below, the effective temperature becomes large. 


\begin{figure}[t]\centering
\includegraphics[scale=0.75]{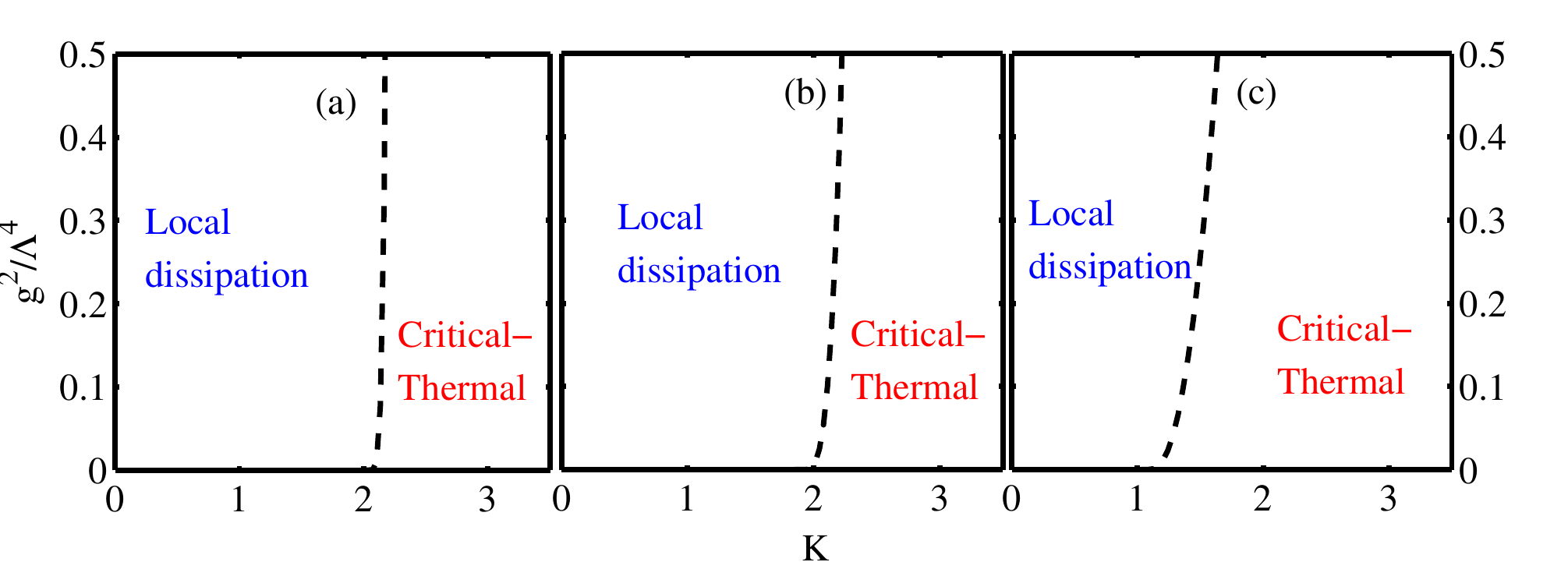}\centering
\caption{Crossover line (\ref{eq:cross3}) as function of the Luttinger parameter $K$ and the commensurate lattice strength $g/\L^2$, for different values of the noise strength $\pi^{-2} F/\eta$: (a) 0.01; (b) 0.1; (c) 1.0. In the ``local dissipative'' regime the effective dissipation is diverging and the effective temperature is small, while in the``critical thermal'' regime the effective temperature is large and the dissipation small. A critical behavior is still observable in the short-times dynamics}\label{fig:onedim}
\end{figure}

\section{Summary and discussion}\label{sec:conclusion}

In this paper we investigated the physics of low dimensional systems driven out of equilibrium by external noise sources. We considered two specific models which in equilibrium undergo interesting phase transitions: (1) a resistively shunted Josephson junction and (2) a one-dimensional quantum liquid in a commensurate lattice potential. The basic question underlying our study is whether non equilibrium systems can exhibit universal physics. We have recently shown\cite{us-nature} that such systems,
in the absence of non-linear coupling between the modes,
establish a critical steady state when driven by $1/f$ noise. Here we used this scale invariant steady state as a starting point for a controlled renormalization group analysis to treat the effect of the non linear couplings and thereby identify universal aspects of the non equilibrium behavior.  The perturbative RG scheme is formulated within the Keldysh path integral framework.

The scaling equation at the first order in the coupling constant indicates a transition between different regimes where the scaling dimension of
a relevant operator controlling the physics of the system (Josephson coupling for the noisy Josephson junction, periodic potential of the lattice for the 1D quantum liquid) changes sign. This occurs as a direct extension of the quantum phase transitions that would exist in the corresponding system
in equilibrium. Here however the scaling dimension can be tuned by varying the external noise strength and not only intrinsic parameters (i.e. shunt resistance or Luttinger parameter). At this level the RG analysis is consistent with the scaling argument we gave in Ref.~\onlinecite{us-nature}.
It is important to note that at this order, although the noise is weakening the various correlation functions of the system, it is acting in a very different way than a temperature, and in particular preserves the power-law decay of the various correlations in the system.

The back action of the nonlinear coupling on the steady state correlations is only given by the RG equations at the second order in the coupling.
We find that generically this leads to the emergence of an effective temperature and effective dissipation in the steady state. In the case of the Josephson junction both effects are perturbatively small, allowing the observation of a universal crossover between an insulating and superconducting regimes, controlled by the above non-equilibrium quantum critical state. Using the RG approach, we obtain analytic expressions for the shape of the crossover, summarized graphically in Fig.~\ref{fig:phdia}. The RG approach allows us to compute the physical properties of each regime, such as the IV characteristic of the junction shown in figures \ref{fig:IVcurve1bis} and \ref{fig:IVcurve2bis}. In general, the IV curves show an algebraic behavior, determined by the underlying critical state, followed by a thermal Ohmic behavior, at voltages lower than the effective temperature. 
		
In the one dimensional case, we considered two types of perturbations around the critical state. The first type, a local impurity, can be mapped into the noisy shunted Josephson junction described above: the one dimensional system plays the role of the resistor and the impurity plays the role of the Josephson coupling. The second type of perturbation, a commensurate lattice, leads to new physics. Despite the different initial set-up, the RG flow is equivalent to the one dimensional quench considered in Ref.~\onlinesquare{GiamarchiMitra}. The resulting non-equilibrium phase diagram shows two distinct regime: a thermal regime and a dissipative  regime. In the thermal regime, the lattice is irrelevant at first order. However at second order in $g$ the coupling of modes at all scales induced by the lattice acts to generate fluctuations and effective dissipation, both proportional to $g^2$. This implies a fluctuation dissipation relation with an effective temperature of the order of the cutoff, from which the coupling $g$ drops out. This seemingly paradoxical result is coming from an inversion of two limits, namely the fact that the coupling to the noise can be small and the fact that one observes the steady state of the system at infinitely long time, allowing a large amount of energy to be pumped into the system under the form of a temperature. However, as the effective dissipation is small, the thermalization rate of the system is vary small, thus allowing to observe transient non-equilibrium phenomena for relatively long times. Clearly, the complete study of the dynamics of the system, in particular at transient times goes beyond the scope of this paper and of the perturbative RG analysis, and deserves further study.

\section{Acknowledgment}
We thank S. Huber,  D. Huse, A. Mitra and A. Rosch for stimulating discussions. This research was supported in part by the US israel binational science foundation (EA and ED), the Israel Science foundation (EA), and the Swiss NSF under MaNEP and Division II (TG). EGDT is supported by the Adams Fellowship Program of the Israel Academy of Sciences and Humanities.

\appendix

\section{Explicit calculation of the RG flow coefficients} \label{app:explicitRG}

In this appendix, we show how to define a proper cutoff  and use it to compute the dimensionless pre-factors appearing in the scaling equations (\ref{eq:scaleT}-\ref{eq:scaleF}) and (\ref{eq:scaleLL1}-\ref{eq:scaleLL4}).
Specifically, we need to obtain explicit expressions for the correlation ${\mathcal K}(x,t)$ and response ${\mathcal R}(x,t)$ of the
exponential operator $e^{i\phi}$ and then use these expressions in order to compute the integrals in
(\ref{eq:I0-general}-\ref{eq:I1-general}) and (\ref{eq:Ieta-general}) and (\ref{eq:Iu-general})

\subsection{Noisy Josephson junction}
\label{app:josephson}

\begin{figure}[t]
\includegraphics[scale=0.7]{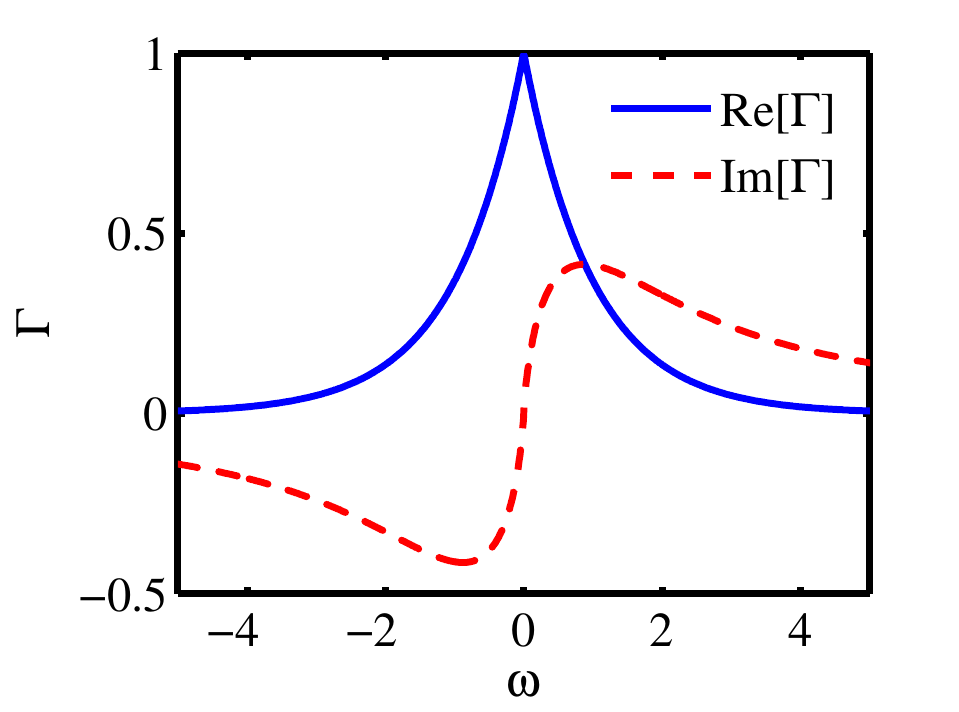}\centering
\caption{Real and imaginary part of the cutoff function (\ref{eq:cutoff}). The real part of this function equals to $e^{-|\w|/\L}$. In addition, the cutoff function has a non trivial imaginary part which allows to preserve causality.}
\label{fig:cutoff}
\end{figure}

The simplest choice for the cutoff would be a step function in frequency, $\Gamma(\w/\L)=1-\Theta(w/\L)$. However, as we noted earlier, this choice is not valid because it violates the causality of the retarded Green's function $G_{R,\L}(\w) = G_{R,\L_0}(\w)\G_{\L}(\w)$. Instead we choose the following cutoff function
\bea
\G\left(\frac{\w}{\L}\right) &=&\frac{1}{\pi} \int d \nu \frac{-i\w}{(\nu-\w-i0^+)(\nu+\w+i0^+)}e^{-|\nu|/\L}\nn\\
&=&e^{-|\w|/\L} + \frac{2i}{\pi}\left[-{\rm CoshIntegral}\left(\frac\w\L\right){\rm Sinh}\left(\frac\w\L\right) +{\rm SinhIntegral}\left(\frac\w\L\right){\rm Cosh}\left(\frac\w\L\right)\right]\label{eq:cutoff}
\eea
Eq. (\ref{eq:cutoff}) explicitly preserves causality and, as shown in Fig.~\ref{fig:cutoff}, satisfies the defining conditions of a cutoff $\Gamma_\L(0)=1$ and $\Gamma(\infty)=0$.

This specific choice allows us to obtain simple expressions for the Green's functions, $G_{K,\L}(t) = \av{\phi(t)\phi(0)}$ and $G_{R,\L}(t) = \av{\hat\phi(0)\phi(t)}$. For convenience, we first compute the derivatives of these functions with respect to $\L$. Assuming zero temperature ($T=0$) we obtain:
\bea
{\partial G_{K,\L}(t)\over \partial \L} &=& i\frac{R}{R_Q}\left(1+{\bar F}\right){\partial\over \partial\L} \int d\w~e^{i\w t}\frac1{\w}\left[\G\left(\frac{\w}{\L}\right)+\G\left(-\frac{\w}{\L}\right)\right] \\
&=& 
i\frac{R}{R_Q}\left(1+{\bar F}\right) \frac1{\L^2}\int_{-\infty}^{\infty}d \nu e^{-|\nu|/\L}\left(e^{i \nu t}+e^{-i \nu t}\right)
= 2i\frac{R}{R_Q}\left(1+{\bar F}\right) \frac{-2/\L}{1+\L^2 t^2}\label{eq:dddC}\\
{\partial G_{R,\L}(t)\over \partial\L} &=& \frac{R}{R_Q}{\partial\over \partial\L} \int d\w~e^{i\w t}\G\left(\frac{\w}{\L}\right)\frac1{i \w t} 
\\
&=& \frac{R}{R_Q} \frac1{\L^2}\int_{-\infty}^{\infty} (u~dq) e^{-u|q|/\L}\left(e^{i q t}-e^{-i q t}\right) =  2\frac{R}{R_Q}\Theta(t)\left(1+{\bar F}\right) \frac{-2t}{1+\L^2 t^2}\label{eq:dddR}
\eea
Here $\Theta(t)$ is the Heaviside step function. We can now integrate with respect to $\L$ and obtain:
\bea
G_{K,\L}(t) &=& G_{K,\L}(0) + 2i\frac{R}{R_Q}(1+{\bar F})\log\left(1+\L^2 t^2\right) \label{eq:Ccritical}\\
G_{R,\L}(t) &=& 2\Theta(t)\frac{R}{R_Q}\arctan\left(\frac{t}{\L}\right) \label{eq:Rcritical}
\eea

The correlation and response functions of the exponential operator $e^{i\theta}$ are in turn simple functions of the Green's functions $G_{K,\L}$ and $G_{R,L}$:
\bea {\mathcal K}(t) &=&\av{e^{i\theta(t)+i\hat\theta(t)}\left(e^{-i\theta(0)-i\hat\theta(0)}+e^{i\theta(0)-i\hat\theta(0)}\right)} \\
&=&e^{-\half\av{\left(\theta(t)+\hat\theta(t)-\theta(0)-\hat\theta(0)\right)^2}}+e^{-\half\av{\left(\theta(t)+\hat\theta(t)-\theta(0)-\hat\theta(0)\right)^2}}\\
&=&2\cos(G_{R,\L}(t')) e^{i\left(G_{K,\L}(t)-G_{K,\L}(0)\right)}\\
{\mathcal R}(t) &=&\av{e^{i\theta(t)+i\hat\theta(t)}\left(e^{-i\theta(0)-i\hat\theta(0)}-e^{i\theta(0)-i\hat\theta(0)}\right)} \\
&=& 2\sin(G_{R,\L}(t')) e^{i\left(G_{K,\L}(t)-G_{K,\L}(0)\right)}\eea
Here we used $G_K=i\av{\theta\theta}$, $G_R=i\av{\hat\theta\theta}$, and $\av{\hat\theta\hat\theta}=0$.

We now use these expressions together with the first order result (\ref{eq:generalcutoff}) to derive
the $I_0$ and $I_1$, defined in (\ref{eq:I0-general}) and (\ref{eq:I1-general})
\bea
I_0 &=&  \left(2\frac{R}{R_Q}(1+{\bar F}) - 1\right)\int_{-\infty}^\infty dt~\cos\left(2\frac{R}{R_Q}{\rm artg}\left(\L t\right)\right)\left(\frac1{1+\L^2 t^2}\right)^{\frac{R}{R_Q}(1+{\bar F})}\label{eq:GK0-1}\\
&=&  \left(2\frac{R}{R_Q}(1+{\bar F}) - 1\right)\int_{-\pi/2}^{\pi/2} dy~\cos\left(2\frac{R}{R_Q}y\right)(\cos y)^{2\frac{R}{R_Q}(1+{\bar F})-2}\label{eq:GK0-2} \\
&=& \frac{4\pi}{2^{2R/R_Q(1+{\bar F})}} \frac1{B\left(\frac{R}{R_Q}{\bar F},\frac{R}{R_Q}(2+{\bar F})\right)}\label{eq:I0} \\
I_1 &=& \left(2\frac{R}{R_Q}(1+{\bar F}) - 2\right) \int_{-\infty}^\infty dt~t~\sin\left(2\frac{R}{R_Q}{\rm artg}\left(\L t\right)\right)\left(\frac1{1+\L^2 t^2}\right)^{\frac{R}{R_Q}(1+{\bar F})}\\
&=&  2 \left(2\frac{R}{R_Q}(1+{\bar F}) - 2\right) \int_{0}^{\pi/2} dy~\sin\left(2\frac{R}{R_Q}y\right)(\cos y)^{2\frac{R}{R_Q}(1+{\bar F})-2}\tan(y) \\
&=& 4\frac{R}{R_Q} \int_0^{\pi/2} dy~\cos\left(2\frac{R}{R_Q}y\right)(\cos y)^{2\frac{R}{R_Q}(1+{\bar F})-2}\label{eq:GR1-1}\\
&=&  \frac{4\pi R/R_Q}{2^{2R/R_Q(1+{\bar F})} \left(2\frac{R}{R_Q}(1+{\bar F})-1\right) B\left(\frac{R}{R_Q}{\bar F},\frac{R}{R_Q}(2+{\bar F})\right)} \label{eq:I1}
\eea
Here $B(x,y)=\Gamma(x+y)/\Gamma(x)\Gamma(y)$ is the Bigamma function. This function diverges if $x=0$ and $y\neq0$, implying that both $I_0$ and $I_1$ vanish at equilibrium (${\bar F}=0$). In addition, we note that eq. (\ref{eq:I1}) is valid only for $2\frac{R}{R_Q}(1+{\bar F})>1$, while for $2\frac{R}{R_Q}(1+{\bar F})<1$ the integral diverges.

\subsection{One dimensional system}
\label{app:1d}
We now move to the one dimensional system. Here the natural choice of the cutoff is $\Gamma(q,\w)=e^{-u|q|/\L}$.  The correlation and response function of $\phi$ are then given by:
\bea
G_K ( x , t) &-& G_K(0,0) = i\half\av{\left(\phi(x,t)-\phi(0,0)\right)^2}\nn\\
&=&i \frac{K}4 \int (u ~dq)~d\w~\left(1-e^{i q \delta x- i\w \delta t}\right) e^{-u|q|/\L}\left(1+ \frac{F}{\pi^2\eta}\frac{q^2}{\w^2}\right) \delta(\w^2- u^2q^2)\nn\\
&=&i \frac{K}{4}\left(1+ {\bar F}\right) \left[\log\left(|1 + \L( x/u - t)|\right)+\log\left(|1 + \L( x/u + t)|\right)\right]\label{eq:Ccritical2}\\
G_R( x, t) &=& i \av{\hat\phi(x,t)\phi(0,0) }\newline = \lim_{\eta\to 0} \int (u ~dq)~d\w~e^{i q  x - i \w t}~e^{-u|q|/\L}\frac{1}{\w^2-u^2q^2-i\eta\w}\\
&=& \frac{K}2 \Theta(t)\left[{\rm artg}(\L( x/u + t))+{\rm artg}(\L(x/u - t))\right]\label{eq:Rcritical2}
\eea
Here we have neglected the effects of both the dissipation and the temperature ($\eta=T=0^+$) and ${\bar F}$ is defined in (\ref{eq:F02})

Substituting these expressions in (\ref{eq:IT-general}-\ref{eq:Iu-general}), we find:
\bea
I_T &=& (2\tilde K-2)\int_{-\infty}^\infty dx \int_{-\infty}^\infty dt \cos\left(K{\rm artan}(x+t) - K{\rm artan} (x-t)\right)\left(\frac1{1+(x+t)^2}\right)^{\tilde K/2}\left(\frac1{1+(x-t)^2}\right)^{\tilde K/2}\\
&=&  4(\tilde K -1)\int_0^\infty du~  \cos\left(K{\rm artan}(u)\right)\left(\frac1{1+u^2}\right)^{\tilde K/2}\int_0^\infty dv~\cos\left(K{\rm artan}(v)\right) \left(\frac1{1+v^2}\right)^{\tilde K/2}\\
&=& (\tilde K -1)\left[\frac{4\pi}{2^{\tilde K}}\frac1{(\tilde K -1)B\left(\frac{\tilde K + K}2,\frac{\tilde K - K}2\right)}\right]^2\label{eq:IT}\\
I_\eta &=& \left(2\tilde K - 3\right)\int_{-\infty}^\infty dx \int_0^\infty dt ~t~ \sin\left(K{\rm artan}(x+t) - K{\rm artan} (x-t)\right)\left(\frac1{1+(x+t)^2}\right)^{\tilde K/2}\left(\frac1{1+(x-t)^2}\right)^{\tilde K/2}\\
&=& 4\left(\tilde K -\frac{3}2\right)\int_0^\infty du~  \cos\left(K{\rm artan}(u)\right)\left(\frac1{1+u^2}\right)^{\tilde K/2}\int_0^\infty dv~v~\sin\left(K{\rm artan}(v)\right) \left(\frac1{1+v^2}\right)^{\tilde K/2}\\
&=&4\left(\tilde K -\frac32\right)\cos\left(K{\rm artan}(u)\right)\left(\frac1{1+u^2}\right)^{\tilde K/2}\int_0^\infty~\frac{K}{\tilde K/2 - 1}\int_0^\infty dv~\cos\left(K{\rm artan}(v)\right) \left(\frac1{1+v^2}\right)^{\tilde K/2}\\
&=&\frac{\left(\tilde K -\frac32\right)}{2\left(\tilde K - 1\right)\left(\tilde K -2\right)}I_T\label{eq:Ieta}\\
I_K &=& \left(2\tilde K - 4\right)\int_{-\infty}^\infty dx \int_0^\infty dt~\left(x^2 + t^2~ \sin\left(K\arctan(x+t) -K\arctan(x-t)\right)\right)\left(\frac1{1+(x+t)^2}\cdot\frac1{1+(x-t)^2}\right)^{\tilde K\over 2}\label{eq:IK}\\
I_u &=& \left(2\tilde K - 4\right)\int_{-\infty}^\infty dx \int_0^\infty dt~\left(x^2-t^2~ \sin\left(K\arctan(x+t) - K\arctan(x-t)\right)\right)\left(\frac1{1+(x+t)^2}\right)^{\tilde K/2}\left(\frac1{1+(x-t)^2}\right)^{\tilde K/2}\label{eq:Iu}
\eea

In the absence of the noise ($F/\eta=0 \Rightarrow \tilde K=K$), both $I_T$ and $I_\eta$ go to zero. As expected at equilibrium, the non-linear coupling does not lead to a renormalization of the temperature or of the dissipation. Also note that, for any finite noise strength, $I_\eta$ diverges at $\tilde K =2$, while $I_T$ remains finite at this point. This observation will have dramatical consequences on the properties of the expected non-equilibrium phase transition\cite{us-nature}, located precisely at $\tilde K=2$.

For the integrals $I_K$ and $I_u$ we were not able to find an analytic solution. However, we have numerically computed them and found that, in agreement with the known equilibrium results\cite{giamarchi}, both integrals do not vanish even in the absence of the noise. In addition, we found that both $I_K$ and $I_u$ are finite for any $\tilde K>2$. For details about their behavior around the $\tilde K=2$ point, the reader is referred to Ref.\onlinesquare{GiamarchiMitra-long}.

\section{Duality transfortmation}
\label{app:duality}
In this appendix we review how the duality transformation\cite{Schmidt,FisherZwerger,AltlandSimons} works within the real time Keldysh framework. 
Our starting point is the non-equilibrium Keldysh action (\ref{eq:SJJ}), which we split into a unitary part and a dissipative or non-local part $S = S_{\rm u} + S_{\rm dis}$. The unitary part of the action includes the capacitor $C$ and the Josephson coupling $g$, while the non local terms include the resistor and the external noise.

In the strong coupling limit ($g\gg \L$) the unitary action is large justifying the use of a Saddle point approximation for the phase dynamics of the field $\t$. This leads to two independent equations of motion for the fields $\phi_\pm(t)$ on the forward and backward branches of the Keldysh contour
\be\frac{\d S_{\rm Re}}{\d\phi_\pm} = (RC)\partial^2_t \phi_\pm(t) - g \sin(\phi_\pm(t)) = 0\label{eq:saddle}\ee
Equation (\ref{eq:saddle}) has instanton solutions $\phi(t)=2\pi \e_s f(t-t_0)$, where $\e_s=\pm1$ and $f(t)$ is a step function broadened by a short time-scale $\tau=\sqrt{C/g}$. The instanton solution describes a jump of the phase (phase slip) by $\pm 2\pi$  between minima of the cosine potential. A more general saddle point solution can include any number $n_+$ of instantons in the forward and $n_-$ instantons in the backward branches of the Keldysh contour. Furthermore, in the strong coupling limit
phase slips are rare, in the sense that the average time between them is much larger than the instanton time $\tau$. In this case we can take a general saddle point solution to be a sum of independent instantons and at the same time regard each instanton as a step function
\be
\phi_\pm(t) = \sum_{i=1}^{n_{\pm}} \e_{\pm,i} \Theta(t-t_{\pm,i}) \label{eq:phidual}\ee
Here $t_{\pm, i}$ is the time of the instanton.
Plugging in these solutions into the Keldysh action, the path integral is converted to a sum over all possible instanton configurations on the two branches
\be
\int\mathcal{D}\phi_+\mathcal{D}\phi_-~ e^{i S[\phi_+,\phi_-]} = \sum_{\alpha=\pm}\sum_{c_\alpha}\frac{1}{n_\a !} \exp\left( i\a n_\a S_{\rm ins}+ iS_{\rm dis}[c_+,c_-]\right).\label{eq:Ssum}
\ee
Here $c_\pm=\{n_\pm,\e_{\pm, i} , t_{\pm,i}\}$ denotes an instanton configuaration
on the forward/backward branch  and $S_{\rm ins}$ is the action of a single instanton.

The non-local part of the action $S_{\rm dis}[c_+,c_-]$ can be written explicitly in terms of the instanton coordinates $t_{i,\a}$ and $\e_{i,\a}=\pm$ by plugging the instanton solutions into the  the quadratic form of the original action action
\be S_{\rm dis} = \sum_{\a,\b=\pm}\int dt~\int dt' \phi_\a(t) \left(G^{-1}_{\rm dis}\right)_{\a\b}(t-t') \phi_\b(t')= \sum_{\a,\b}\sum_{i,j=1}^{n_{\pm}}\e_{\a,i}\e_{\b,j} \int_{-\infty}^{t_{\a,i}} dt_1 \int_{-\infty}^{t_{\b,j}} dt_2  \left(G^{-1}_{\rm Im}\right)_{\a\b}(t_1 - t_2). \label{eq:Simag}\ee
This part of the action contributes a pairwise interaction between instantons at different times.

We now introduce the dual field $\theta_\a(t)$ through the identity
\be e^{{i S_{\rm dis}[\e_i,t_i]}} = \int \mathcal{D}\theta \exp\left( i \sum_{\a,\b}\a\b\int d\w~\w^2 \theta^\a(-\w) G_{\rm dis}(\w)_{\a\b}\theta^\b(\w) + \sum_\a \sum_{i=1}^{n_\a} \a\e_{\a,i}\theta_\a(t_{i})\right) \label{eq:Simag2}\ee
Substituting this back into (\ref{eq:Ssum}) we can perform the summation over $n_\a$ as follows
\bea
\sum_{n_\a}\frac{1}{n_\a !} \exp\left(i n_\a S_{\rm sol}+  i\sum_{i=1}^{n_\a} \e_{\a,i}\theta_\a(t_{i})\right)&=&\sum_{n} \frac{1}{(n)!} \left(\sum_{\e,t} e^{i S_{\rm sol} + i\e \theta_\a(t)}\right)^n \\
= \exp\left({\sum_{t} e^{i S_{\rm sol}}\cos(\theta_\a(t))}\right) &\equiv& \exp\left(i~g_{\rm dual} \int dt \left(\cos(\theta_\a(t))\right)\right)\label{eq:Scos}
\eea
Here we used the time scale $RC$ to transform the sum over times into an integral. The phase slip fugacity $g_{\rm dual} = (1/RC) e^{i S_{\rm sol}}$ be estimated using the WKB approximation\cite{Schmidt}
to be  $g_{\rm dual} \approx (1/RC) e^{-\sqrt{g RC}}$.

Finally, putting (\ref{eq:Scos}) together with (\ref{eq:Simag2}), we obtain the dual action
\be S = \sum_{\a,\b}\int d\w~ \theta_{\a}(-\w)(G^{-1}_{\rm dual})^{\a\b}\theta_{\b}(\w) + g_{\rm dual} \int dt \cos(\theta_+)-\cos(\theta_-)\label{eq:Sdual}\ee
We have defined $G^{-1}_{\rm dual}(\w)_{\a\b}= \a\b~ \w^2~ G_{\rm dis}(\w)_{\a,\b}$.

To compare the dual model (\ref{eq:Sdual}) with the original one (\ref{eq:SJJ}), it is useful to express $G^{-1}_{\rm dual}$ in terms of its  classical and quantum components:
\be \left(\ba{c c}0 & G^{-1,R}_{\rm dual} \\ G^{-1,A}_{\rm dual} & G^{-1,K}_{\rm dual} \ea\right)= \w^2 \left(\ba{c c}0 & G^{A}_{\rm Im} \\ G^{R}_{\rm Im} & - G^{K}_{\rm Im} \ea\right)\ee
Using (\ref{eq:SJJ}) we obtain:
\be G^{-1}_{\rm dual} = \left(\ba{c c}0 & i\frac{R}{R_Q}\w \\ -i\frac{R}{R_Q}\w & -2i \frac{R^2}{R^2_Q}\left(\frac{R_Q}{R}\w{\rm cotgh(\w/2T)} + F\w^{2-\a}\right) \ea\right)\label{eq:S0dual}\ee
 By inspection we see that the dual action is identical to the original one, up to the duality transformation $R/R_Q\to R_Q/R$, $F\to (R/R_Q)^2 F$, $T\to T$, and $g\to g_{\rm dual}$. In this sense the shunted Josephson junction is self dual even in the presence of the external noise.

In the above derivation, we implicitly assumed that the junction is disconnected from external sources apart from the noise source. However to derive the current voltage characteristics we need to embed the junction in a circuit with applied current or voltage bias and to generalize the duality transformation to this case.The question is how the circuit structure transforms under the duality transformation.

Consider the circuit shown in Figure \ref{fig:SJJdual}(a), in which the noise-driven shunted Josephson junction is biased by a DC current. The current can be enforced by adding the term $S_I = I \int dt ~ \phi(t)$ to the Keldysh action (\ref{eq:SJJ}), which transforms the periodic cosine potential into a wash-board potential. If the current is small compared to $g$, we can still substitute the soliton solutions obtained above into the action. Doing this for the new term $S_I$ we obtain $S_I =2\pi I \sum_{\a=\pm}\sum_i \a\e_{i\a} t_{i\a}$. Introducing the dual field $\theta$ as before and summing over $n$, we get
\be S = \sum_{\a,\b}\int d\w~ \w^2 \theta_{\a}(-\w)(G^{-1}_{\rm dual})_{\a,\b}\theta_{\b}(\w) + g \int dt \cos(\theta_+(t)+ R_Q It)-\cos(\theta_-(t)+  R_Q It)\label{eq:Sdual2}\ee
(here we work in the units for which $\hbar=1$ and $R_Q = h/(2e)^2 = 2\pi$). Equation (\ref{eq:Sdual2}) shows that a current $I$ acts in the dual circuit as a voltage bias  $V_{\rm dual} = R_Q I$, applied only to the Josephson junction (and not to the resistor). In this sense, the dual model corresponds to a circuit in which the Josephson junction is connected in series to the resistor (See Figure \ref{fig:SJJdual} (c)).

We will now show that the duality transformation maps the voltage across the original circuit into the current of the dual circuit. The voltage can be obtained formally by taking the derivative of the Keldysh path integral with respect to separate current sources $I_+$ and $I_-$ acting in the forward and backward directions and setting $I_+=I_-=I$ in the end:
\bea \av{V(t)} &=& \partial_t \av{\theta(t)} = \partial_t \int d\theta~\left.\left(\frac{d}{d I_+(t)}-\frac{d}{d I_-(t)}\right)e^{i S[I_+,I_-]}\right|_{I_\pm=I} \nn\\
 &\equiv& R_Q ~ g_{\rm dual} \partial_t \int d\theta ~\left[\sin(\theta_+ + R_Q I t)+\sin(\theta_- + R_Q I t)\right]~t~e^{i S_{\rm dual}}\label{eq:Idual}\eea
As the junction is in a steady state, $\av{V(t)}$ is independent on $t$ and we can take for example $t=0$:
\be \av{V}= R_Q~g_{\rm dual}\int d\theta ~\left[\sin(\theta_+)+\sin(\theta_-)\right]~e^{i S_{\rm dual}} \equiv R_Q\av{I^s_{\rm dual}}\label{eq:Idual2}\ee
We see from the last expression that $\av{V(t)}=R_Q I^s_{\rm dual}$ by definition of the super-current trough the dual junction. Note that since the Junction is in steady state the expectation value cannot be time dependent and we might as well set $t=0$ in Eq. (\ref{eq:Idual2}).

Equations (\ref{eq:Sdual2}) and (\ref{eq:Idual2}) can be used to evaluate the IV characterstic of the original circuit, in its strong coupling regime. (The calculation follows the same steps used to compute the current across the junction in the weak coupling regime.) In the strong coupling regime $g\gg \L \Rightarrow g_{\rm dual}\ll \L$ and we can apply the first order perturbation theory in $g_{\rm dual}$ to find
\be\av{V} = i~g_{\rm dual}^2\int_{-\infty}^0 dt' \av{\left[\cos(\theta_+(t')+R_Q I t')-\cos(\theta_-(t')+R_Q It')\right]\left[\sin(\theta_+(0))+\sin(\theta_-(0))\right]}_{0}\label{eq:Idual1}\ee
Here $\av{...}_0$ is the average with respect to the quadratic part of the Keldysh action (\ref{eq:Sdual}). Finally, we exploit causality to extend the integral from $t'=0$ to $t'=\infty$ and re-write (\ref{eq:Idual1}) as
\bea\av{V} &=& g_{\rm dual}^2{\rm Im}\left[\int_{-\infty}^\infty dt' e^{i R_Q I t}\av{\left(e^{i\theta_+(t')}-e^{i\theta_-(t')}\right)\left(e^{i\theta_+(0)}+e^{i\theta_-(0)}\right)}_{0}\right]\\
&\equiv& g_{\rm dual}^2 {\rm Im}\left[ {\mathcal R}(\w=R_Q I)\right]
\label{eq:Idual-pert}\eea
Here ${\mathcal R}(\w)$ is the Fourier transform of the response function of the operator $e^{i\theta}$.

\bibliographystyle{apsrev}
\bibliography{noneq}
	
\end{document}